\documentclass[a4paper]{article}

\usepackage{jheppub} 
                     
\usepackage{graphicx,color}
 \usepackage{bm}
   \usepackage{amsmath}
    \usepackage{amssymb}    
     \usepackage{pifont}
\usepackage{tikz}
\usetikzlibrary{decorations.pathmorphing}
\usetikzlibrary{decorations.markings}
\usetikzlibrary{calc}
\usetikzlibrary{math}
\usetikzlibrary{patterns}
\usepackage{braket}
\usepackage{standalone}
\usepackage{slashed}
\usepackage{multirow}
\usepackage{hhline}
\usepackage{colortbl}

\newcommand{\Ds}{\displaystyle}

\newcommand{\nn}{\nonumber}

\newcommand{\ot}{\leftarrow}

\renewcommand{\(}{\left(}
\renewcommand{\)}{\right)}
\renewcommand{\[}{\left[}
\renewcommand{\]}{\right]}
\newcommand{\ta}{\left(}
\newcommand{\tc}{\right)}

\newcommand{\lDer}[1]{\overset{\leftarrow}{#1}\phantom{\,}}

\newcommand{\e}{\epsilon}

\title{Sivers, Boer-Mulders and worm-gear distributions at next-to-leading order}

\author[a]{Felix Rein}
\author[a]{Simone Rodini}
\author[a]{Andreas Sch\"afer}
\author[b]{Alexey Vladimirov}

\affiliation[a]{Institut f\"ur Theoretische Physik, Universit\"at Regensburg, D-93040 Regensburg, Germany}
\affiliation[b]{Dpto. de F\'isica Te\'orica \& IPARCOS, Universidad Complutense de Madrid, E-28040 Madrid, Spain}

\emailAdd{simone.rodini@physik.uni-regensburg.de}
\emailAdd{alexeyvl@ucm.es}

\abstract{
We compute the Sivers, Boer-Mulders, worm-gear (T and L) transverse momentum dependent distributions in terms of twist-two and twist-three collinear distributions in the small-$b$ limit up to next-to-leading order (NLO) in perturbation theory. 
}

\begin{document} 
\allowdisplaybreaks
\maketitle 

\section{Introduction}

The leading power transverse momentum dependent factorization theorem introduces eight quark transverse momentum-dependent distributions (TMDs) \cite{Mulders:1995dh, Boer:1997nt, Boer:1999uu}, which are listed in table \ref{tab:tmds}. Altogether, these eight TMDs provide a comprehensive description of the nucleon's three-dimensional spin-orbital structure in momentum space. Some of these TMDs (primarily the unpolarized ones) are studied very well theoretically and experimentally (for recent developments, see \cite{Scimemi:2019cmh, Bacchetta:2022awv}). However, several of these TMDs are still almost unexplored. This paper is devoted to study the Sivers, Boer-Mulders, worm-gear-T, and worm-gear-L (also known as Kotzinian–Mulders) functions in the limit of small-$b$ (or, equivalently, large transverse momentum) within QCD perturbation theory.

TMDs are nonperturbative functions of two kinematic variables $x$ and $k_T$, being $x$ the collinear momentum-fraction and  $k_T$ the transverse momentum. Equivalently, one can use Fourier transformed TMDs $k_T$-space to position space, labelling the transverse coordinate vector with $b$. In many aspects, the position space definition is advantageous. We use it throughout the work, referring to the distributions depending on $x$ and $b$ as TMDs. Different ranges of $x$ and $b$ correspond to different physical pictures, relevant for different processes. In particular, in the limit of small $b$, TMDs turn into ordinary one-dimensional collinear parton distributions. Schematically, this relation has the form
\begin{eqnarray}\label{into1}
F(x,b)=C(x,\ln(\mu b))\otimes f(x,\mu)+\mathcal{O}(b^2),
\end{eqnarray}
where $F$ is a TMD, $f$ is a collinear distribution, $C$ is a perturbative coefficient function, and $\otimes$ is an integral convolution. The expansion (\ref{into1}) (also known as the ``matching relation''  \cite{Collins:2011zzd}) follows from the operator product expansion (OPE) and can be derived systematically order-by-order in the coupling constant and  powers of $b^2$ \cite{Moos:2020wvd}. 

Small-$b$ expansions for TMDs have been intensively studied during the last decade. Naturally, the main efforts were devoted to the unpolarized distribution $f_1$, for which the coefficient function is known at next-to-next-to-next-to-leading order (N$^3$LO) in the QCD coupling constant \cite{Ebert:2020qef,Luo:2020epw}. For the other distributions, the analysis is less developed. So, the transversity $h_1$ and linearly-polarized gluon TMD $h^g_1$ are known up to NNLO \cite{Gutierrez-Reyes:2018iod, Gutierrez-Reyes:2019rug}. The helicity $g_1$ is known at NLO \cite{Gutierrez-Reyes:2017glx, Buffing:2017mqm, Bacchetta:2013pqa}. All these TMDs are special because their small-$b$ asymptotic contains only collinear distributions of twist-two. Therefore, their computation is relatively straightforward and can be done with standard techniques. However, the majority of TMDs match collinear distributions of higher twists, making their study more cumbersome. Thus, for Boer-Mulders $h_1^\perp$, worm-gear-T $g_{1T}$, and worm-gear-L $h_{1L}^\perp$ the small-$b$ expansion is known only 
at LO 
\cite{Kanazawa:2015ajw, Scimemi:2018mmi, Meissner:2007rx,Boer:2003cm} with some partial results known at NLO \cite{Ji:2006ub,Zhou:2009jm,Dai:2014ala}, and for the Sivers function $f_{1T}^\perp$ at NLO \cite{Scimemi:2019gge}. The pretzelosity distribution $h_{1T}^\perp$ differs from other TMDs. Its leading term is given by a twist-four operator, while matching is only known for the twist-three part \cite{Moos:2020wvd}. In table \ref{tab:tmds} we indicate the twists of collinear distributions that appear as the leading-power term in eqn.~(\ref{into1}).

\begin{table}[tb]
\begin{center}
\begin{tabular}{|c||c|c|c|}
\hline
& U & H & T
\\\hline
U & $f_1$ (tw2) & & \cellcolor{blue!25} $h_1^\perp$ (tw3)
\\\hline
L &  & $g_1$ (tw2) & \cellcolor{blue!25}$h_{1L}^\perp$ (tw2 \& tw3) 
\\\hline
\multirow{2}{*}{T} & \cellcolor{blue!25} $f_{1T}^\perp$ (tw3) & \cellcolor{blue!25} $g_{1T}$ (tw2 \& tw3) & $h_1$ (tw2) 
\\ &\cellcolor{blue!25}&\cellcolor{blue!25} & $h_{1T}^\perp$ (tw3 \& tw4)
\\\hline
\end{tabular}
\caption{\label{tab:tmds} Quark TMDs sorted with respect to polarization properties of both the operator (columns) and the hadron (rows). The labels U, H, L, and T, are for the unpolarized, helicity, longitudinal, and transverse polarizations. In brackets, we indicate the twist of collinear distributions to which TMDs match at small-$b$. The blue color highlights TMDs that are investigated in this work.}
\end{center}
\end{table}
 
The usage of matching relations is essential for practical applications. It allows incorporating the already-known parton distribution functions into TMDs, essentially increasing the predictive power of the formalism. In fact, all modern phenomenological extractions of TMDs are based on these relations (see f.i. \cite{Bacchetta:2017gcc, Scimemi:2017etj, Scimemi:2019cmh, Bacchetta:2022awv, Cammarota:2020qcw, Echevarria:2020hpy, Bury:2022czx}). 
The twist-two part of the matching relation (the so-called Wandzura-Wilczek-like (WW-like) approximation) is supposed to work fairly well for many cases \cite{Cammarota:2020qcw, Bhattacharya:2021twu}. Also, matching relations can be inverted and used to determine collinear distributions from TMDs. For example, the knowledge of Sivers function provides an essential constraint on the Qiu-Sterman twist-three distribution \cite{Bury:2021sue, Bury:2020vhj}. Finally, the relation (\ref{into1}) links TMD factorization theorem to resummation approach \cite{Collins:1981uk}, which is vital for the description of the high-energy data. In all these cases, it is critical to employ at least NLO expressions to fix the scaling properties of distributions.

This contribution aims to close the remaining gap in the theoretical description of polarized TMDs and compute the small-$b$ expansion for TMDs with leading twist-three contributions at NLO. This includes the Sivers, Boer-Mulders, worm-gear-T, and worm-gear-L functions, highlighted in table \ref{tab:tmds}.

There are several approaches to compute higher-twist contributions to the small-$b$ asymptotics of TMDs \cite{Kanazawa:2015ajw, Sun:2013hua, Dai:2014ala, Scimemi:2018mmi, Scimemi:2019gge, Moos:2020wvd}. Among them, the most practical for the present case is the method used in ref.~\cite{Scimemi:2019gge}, i.e. the background-field method with collinear counting. This method is a generalization of the classical approach to deep-inelastic scattering (DIS) \cite{Balitsky:1987bk}. It has been used recently for many higher-twist computations including quasi- and pseudo-distributions \cite{Braun:2021aon, Braun:2021gvv}, leading and sub-leading power TMDs \cite{Scimemi:2019gge,  Vladimirov:2021hdn, Rodini:2022wki}. In many aspects, the work presented here is the straightforward generalization of the computation performed in ref. \cite{Scimemi:2019gge} for different polarizations (we also recompute the Sivers function as a cross-check). Therefore, we do not provide a detailed description of the method, which can be found in the refs. \cite{Scimemi:2019gge, Braun:2021aon}  together with computational examples. Instead, we provide a general discussion, emphasizing the present case's particularities, and present the final expression.

The paper is structured as follows. In section \ref{sec:definitions}, we collect the definitions of TMDs and collinear distributions -- the main subjects of the present work. In section \ref{sec:details}, we provide the essential details on the computation method (referring, for an extended discussion, to \cite{Scimemi:2019gge,Braun:2021aon}). The generalization of $\gamma^5$ in $d$ dimensions and the definition of gluon correlator)  are described in more detail in sections \ref{sec:gamma5} and \ref{sec:FDF}, respectively. In section \ref{sec:results}, we present NLO expressions for Sivers, Boer-Mulders, and worm-gear functions in momentum-fraction space. The position space expressions (split into contributions from the different diagrams) are given in appendix \ref{app:pos}. In appendix \ref{app:evol} are collected the expressions for the twist-three evolution kernels used as cross-check of our computation. 

\section{Definition of distributions}
\label{sec:definitions}

In this work, we deal with many parton distributions. For clarity, we collect their definition and important properties in this section.

\subsection{TMD distributions}

The quark TMDs are defined for the Drell-Yan process, taken as an example, by the following matrix element
\begin{eqnarray}\label{def:TMD-position}
\Phi^{[\Gamma]}(x,b)=\frac{1}{2}\int \frac{d z}{2\pi} e^{-ix zp^+}
\langle p,S|\bar q(zn+b)[zn+b,-\infty n+b]\Gamma [-\infty n,0]q(0)|p,S\rangle,
\end{eqnarray}
where $n$ is the light-like vector ($n^2=0$) associated with the large component of the hadron momentum $p$, $b$ is the vector tranverse to the $(p,n)$ plane, and $\Gamma$ is a Dirac matrix. $[x,y]$ is the straight Wilson lines from $x$ to $y$,
\begin{eqnarray}
[a_1n+b,a_2n+b]=P\exp\(ig\int_{a_2}^{a_1}d\sigma n^\mu A_\mu(\sigma n+b)\).
\end{eqnarray}

The standard parameterization of the matrix element (\ref{def:TMD-position}) can be found in ref.~\cite{Mulders:1995dh}. It reads
\begin{eqnarray}\label{def:TMDs:1:g+}
\Phi^{[\gamma^+]}(x,b)&=&f_1(x,b)+i\epsilon^{\mu\nu}_T b_\mu s_{T\nu}M f_{1T}^\perp(x,b),
\\\label{def:TMDs:1:g+5}
\Phi^{[\gamma^+\gamma^5]}(x,b)&=&\lambda g_{1}(x,b)+i(b \cdot s_T)M g_{1T}(x,b),
\\\label{def:TMDs:1:s+}
\Phi^{[i\sigma^{\alpha+}\gamma^5]}(x,b)&=&s_T^\alpha h_{1}(x,b)-i\lambda b^\alpha M h_{1L}^\perp(x,b)
\\\nn && +i\epsilon^{\alpha\mu}b_\mu M h_1^\perp(x,b)-\frac{M^2 b^2}{2}\(\frac{g_T^{\alpha\mu}}{2}-\frac{b^\alpha b^\mu}{b^2}\)s_{T\mu}h_{1T}^\perp(x,b),
\end{eqnarray}
where $b^2<0$. Here,
\begin{eqnarray}
g_T^{\mu\nu}=g^{\mu\nu}-n^\mu \bar n^\nu-\bar n^\mu n^\nu,\qquad \epsilon^{\mu\nu}_T=\bar n_\alpha n_\beta \epsilon^{\alpha\beta\mu\nu}=\epsilon^{-+\mu\nu},
\end{eqnarray}
where $\bar n^\mu$ is light-cone vector ($\bar n^2=0$) associated with the small-component of the hadron momentum, i.e. $p^\mu=p^+ \bar n^\mu+M^2 n^\mu/(2p^+)$ with $p^+=(n\cdot p)$. The relative normalization is $(n\cdot\bar n)=1$.  The Levi-Civita tensor and $\gamma^5$-matrix are defined in $4$ dimensions as 
\begin{eqnarray}
\epsilon^{0123}=+1,\qquad \gamma^5=-\frac{i}{4!}\epsilon^{\mu\nu\alpha\beta}\gamma_\mu \gamma_\nu\gamma_\alpha\gamma_\beta.
\end{eqnarray}
Consequently, $\epsilon_T^{12}=\epsilon_{T,12}=+1$. 

The variables $\lambda$ and $s_T$ are longitudinal and transverse components of the spin vector
\begin{eqnarray}
s^\mu=\lambda\frac{p^+ \bar n^\mu}{M}-\lambda\frac{n^\mu M}{2p^+}+s_T^\mu,
\end{eqnarray}
where $M$ is the mass of the hadron. This implies $\lambda=M s^+/p^+$.

All TMDs are dimensionless real functions that depend on $b^2$ (the argument $b$ is used for shortness). In this work, we consider only Sivers ($f_{1T}^\perp$), Boer-Mulders ($h_1^\perp$), worm-gear-T ($g_{1T}$) and worm-gear-L ($h_{1L}^\perp$) functions.

The definition (\ref{def:TMD-position}) in a SIDIS-like process has the Wilson line pointing to $+\infty n$ \cite{Boer:2003cm} instead to $-\infty n$. The T-even TMDs (in the present context, these are the worm-gear functions, $g_{1T}$ and $h_{1L}^\perp$) are independent of the direction of the staple contour due to the T-invariance of QCD. They are the same for Drell-Yan-like and SIDIS-like cases. In contrast, the T-odd TMDs (Sivers $f_{1T}^\perp$ and Boer-Mulders $h_{1}^\perp$ functions) dependent on the direction of the staple contour. One has \cite{Collins:2002kn}
\begin{eqnarray}\label{sign-change}
f_{1T}^\perp(x,b)\Big|_{\text{DY}}=-f_{1T}^\perp(x,b)\Big|_{\text{SIDIS}},\qquad
h_{1}^\perp(x,b)\Big|_{\text{DY}}=-h_{1}^\perp(x,b)\Big|_{\text{SIDIS}}.
\end{eqnarray}
Apart of the sign-change the TMDs are identical for both cases. In the following, we assume the DY-like definition, if not specified.

The bare TMDs contain two types of divergences -- ultraviolet and rapidity divergences.  Both types of divergences are multiplicatively renormalizable \cite{Vladimirov:2017ksc}. As a consequence, the renormalized TMD depends on two scales $\mu$ and $\zeta$. These dependencies are described by the evolution equations
\begin{eqnarray}\label{TMD-evol}
\mu^2 \frac{d F(x,b;\mu,\zeta)}{d\mu^2}=\frac{\gamma_F(\mu,\zeta)}{2}F(x,b;\mu,\zeta),
\qquad
\zeta \frac{d F(x,b;\mu,\zeta)}{d\zeta}=-\mathcal{D}(b,\mu)F(x,b;\mu,\zeta),
\end{eqnarray}
where $F$ is any TMD, $\gamma_F$ is the TMD anomalous dimension, and $\mathcal{D}$ is the Collins-Soper kernel \cite{Collins:1981uk}. At LO, these kernels are \cite{Aybat:2011zv}
\begin{eqnarray}
\gamma_F(\mu,\zeta)=a_s(\mu) \(4C_F\mathbf{l}_\zeta+6C_F\)+\mathcal{O}(a_s^2),
\qquad
\mathcal{D}(b,\mu)=a_s(\mu) 2C_F\mathbf{L}_b+\mathcal{O}(a_s^2,b^2),
\end{eqnarray}
where
\begin{eqnarray}\label{def:logs}
a_s(\mu)=\frac{g^2}{(4\pi)^2},\qquad \mathbf{l}_\zeta=\ln\(\frac{\mu^2}{\zeta}\),\qquad \mathbf{L}_b=\ln\(\frac{(-b^2)\mu^2}{4e^{-2\gamma_E}}\),
\end{eqnarray}
with $g$ being the QCD coupling constant, and $\gamma_E$ is the Euler-Mascheroni constant. In the following text, we often omit the scales $(\mu,\zeta)$ to simplify notation. These scales can be reconstructed from the context.

The relation between momentum and position space TMDs is
\begin{eqnarray}
\Phi^{[\Gamma]}(x,k_T)=\int \frac{d^2b}{(2\pi)}e^{-i(b\cdot k_T)}\Phi^{[\Gamma]}(x,b),
\end{eqnarray}
where $k_T$ is the transverse momentum ($k_T^2<0$). The transformations for individual TMDs can be found in refs. \cite{Boer:2011xd, Scimemi:2018mmi}. The momentum-space definition is less convenient for theoretical computations. Therefore, in the following, we use only position space TMDs.

\subsection{Collinear distributions of twist-two}

The collinear distributions of twist-two are defined as follows (see e.g. \cite{Jaffe:1996zw})
\begin{eqnarray}\label{def:f1-coll}
\langle p,S|\bar q(zn)[zn,0] \gamma^+q(0)|p,S\rangle &=& 2 p^+ \int_{-1}^1 dx e^{i xzp^+}f_1(x),
\\\label{def:g1-coll}
\langle p,S|\bar q(zn)[zn,0] \gamma^+\gamma^5 q(0)|p,S\rangle &=& 2 \lambda p^+ \int_{-1}^1 dx e^{i xzp^+}g_1(x),
\\\label{def:h1-coll}
\langle p,S|\bar q(zn)[zn,0] i\sigma^{\alpha +}\gamma^5 q(0)|p,S\rangle &=& 2 s_T^\alpha p^+ \int_{-1}^1 dx e^{i xzp^+}h_1(x),
\end{eqnarray}
where $\alpha$ is a transverse index. These distributions are known as unpolarized ($f_1$), helicity ($g_1$) and tranversity distributions ($h_1$).  They are defined for $x\in[-1,1]$ and are zero for $|x|>1$. The distributions with negative $x$ are usually interpreted as distributions of antiquarks, 
\begin{eqnarray}\nn
f_1(x)&=&\theta(x)f_{1,q}(x)-\theta(-x)f_{1,\bar q}(-x),
\\\label{definite-flavor}
g_1(x)&=&\theta(x)g_{1,q}(x)+\theta(-x)g_{1,\bar q}(-x),
\\\nn
h_1(x)&=&\theta(x)h_{1,q}(x)-\theta(-x)h_{1,\bar q}(-x).
\end{eqnarray}
In the present work, the unpolarized distribution does not appear, and is presented here only for comparison.

Note that the notation $f_1$, $g_1$ and $h_1$ is the same for TMD distributions and collinear distributions. We distinguish these functions by their arguments, which are $(x,b)$ for TMDs and $(x)$ for collinear distributions.

The gluon collinear distributions are defined as
\begin{eqnarray}\label{def:gluon-coll}
\langle p,S|F_{\mu+}(zn)[zn,0] F_{\nu+}(0)|p,S\rangle &=&(p^+)^2 \int_{-1}^1 dx e^{i xzp^+} \frac{x}{2}\(-g^{\mu\nu}_T f_g(x)-i\epsilon^{\mu\nu}_T \Delta f_g(x)\),
\end{eqnarray}
where $f_g$ and $\Delta f_g$ are unpolarized and helicity gluon distributions. Gluon distributions satisfy the ralation
\begin{eqnarray}
f_g(-x)=-f_g(x),\qquad \Delta f_g(-x)=+\Delta f_g(x).
\end{eqnarray}
In dimensional regularization (with $d=4-2\epsilon$) the definition of gluon distributions (\ref{def:gluon-coll}) is modified and takes the form
\begin{eqnarray}\label{def:gluon-coll-d}
\langle p,S|F_{\mu+}(zn)[zn,0] F_{\nu+}(0)|p,S\rangle &=&(p^+)^2 \int_{-1}^1 dx e^{i xzp^+} \frac{x}{2}\(-\frac{g^{\mu\nu}_Tf_g(x)}{1-\epsilon} -\frac{i\epsilon^{\mu\nu}_T\Delta f_g(x)}{(1-\epsilon)(1-2\epsilon)} \),
\end{eqnarray}
where $\epsilon_T^{\mu\nu}$ is the $d$-dimensional generalized Levi-Civita tensor (see sec. \ref{sec:gamma5}). The $\epsilon$-dependent factors are chosen such that the contraction of the correlator's matrix element with $g_T^{\mu\nu}$ or $\epsilon_T^{\mu\nu}$ yields the same result in any dimension.

The scale-dependence of a twist-two distribution $F$ is given by the DGLAP-type equation
\begin{eqnarray}\label{evol-tw2}
\mu^2 \frac{d F_f(x,\mu)}{d\mu^2} = \sum_{f'}\int_x^1 \frac{dy}{y} P_{f\ot f'}(y) F_{f'}\(\frac{x}{y},\mu\),
\end{eqnarray}
where $f$ labels the partons flavor, and $P$ is the evolution kernel. In this work we need only LO expressions for $P$, which can be found, e.g., in \cite{Jaffe:1996zw}.

\subsection{Collinear distributions of twist-three}

The twist-three distributions parametrize the three-point light-cone operators. The quark-gluon-quark distributions are defined as
\begin{eqnarray}
&& \langle p,S|g\bar q(z_1n)F^{\mu+}(z_2n) \gamma^+q(z_3n)|p,S\rangle 
\\\nn &&
\qquad\qquad
= 2 \epsilon^{\mu\nu}_T s_\nu (p^+)^2 M \int[dx] e^{-ip^+(x_1z_1+x_2z_2+x_3z_3)}T(x_1,x_2,x_3),
\\
&&\langle p,S|g\bar q(z_1n)F^{\mu+}(z_2n) \gamma^+\gamma^5q(z_3n)|p,S\rangle 
\\\nn &&
\qquad\qquad
= 2i s_T^\mu (p^+)^2 M \int[dx] e^{-ip^+(x_1z_1+x_2z_2+x_3z_3)}\Delta T(x_1,x_2,x_3),
\\
&&\langle p,S|g\bar q(z_1n)F^{\mu+}(z_2n) i\sigma^{\nu+}\gamma^5q(z_3n)|p,S\rangle 
\\\nn &&
\qquad\qquad
= 2 (p^+)^2 M \int[dx] e^{-ip^+(x_1z_1+x_2z_2+x_3z_3)}
\(\epsilon^{\mu\nu}_T E(x_1,x_2,x_3)+i\lambda g^{\mu\nu}_T H(x_1,x_2,x_3)\),
\end{eqnarray}
where $F_{\mu\nu}$ is the gluon field-strength tensor, and we have omitted the Wilson links $[z_1n,z_2n]$ and $[z_2n,z_3n]$ for brevity. The integral measure 
\begin{eqnarray}\label{def:[dx]}
\int [dx]=\int_{-1}^1 dx_1 d x_2 dx_3 \delta(x_1+x_2+x_3),
\end{eqnarray}
reflects momentum conservation. 
Note that in the above definitions, by convention, the phase of the exponential has the opposite sign compare to the twist-2 distributions.

The quark-gluon-quark distributions are real-valued functions that satisfy the symmetry relations
\begin{align}\label{def:sym-quark}
T(x_1,x_2,x_3)&=T(-x_3,-x_2,-x_1),&
\qquad
\Delta T(x_1,x_2,x_3)&=-\Delta T(-x_3,-x_2,-x_1),
\\\nn
E(x_1,x_2,x_3)&=E(-x_3,-x_2,-x_1),&
\qquad
H(x_1,x_2,x_3)&=-H(-x_3,-x_2,-x_1).
\end{align}
Often it is convenient to use the following combination
\begin{eqnarray}
S^\pm(x_1,x_2,x_3)=\frac{-T(x_1,x_2,x_3)\pm \Delta T(x_1,x_2,x_3)}{2}.
\end{eqnarray}
In the literature one can find different notations for these distributions \cite{Kang:2008ey,Boer:1997bw,Kanazawa:2000hz,Boer:2003cm,Eguchi:2006mc}. For example, ref. \cite{Kang:2008ey} defines
$\widetilde{\mathcal{T}}_{q,F}(x_3,-x_1)=MT(x_1,-x_1-x_3,x_3)$, and 
$\widetilde{\mathcal{T}}_{\Delta q,F}(x_3,-x_1)=M\Delta T(x_1,-x_1-x_3,x_3)$, and ref. \cite{Scimemi:2018mmi} defines $\delta T_\epsilon=E$ and $\delta T_g=H$. A dictionary between the different notations is provided by ref. \cite{Scimemi:2018mmi}.

For the three-gluon distributions, a standard definition has not yet been established. In the literature, one can find several notation for the parametrization of the same three-gluon correlators \cite{Braun:2000yi, Kang:2008ey, Beppu:2010qn, Scimemi:2019gge}. Here we follow the convention of ref. \cite{Scimemi:2019gge}, in which the three-gluon correlators are parametrized as
\begin{eqnarray}
&& \langle p,S|igf^{ABC}F^{\mu+}_A(z_1n)F^{\nu+}_B(z_2n) F^{\rho+}_C(z_3n)|p,S\rangle 
\\\nn &&
\qquad\qquad
= 
(p^+)^3 M \int[dx] e^{-ip^+(x_1z_1+x_2z_2+x_3z_3)}\sum_{i}t_i^{\mu\nu\rho}F_i^+(x_1,x_2,x_3),
\\
&& \langle p,S|gd^{ABC}F^{\mu+}_A(z_1n)F^{\nu+}_B(z_2n) F^{\rho+}_C(z_3n)|p,S\rangle 
\\\nn &&
\qquad\qquad
= 
(p^+)^3 M \int[dx] e^{-ip^+(x_1z_1+x_2z_2+x_3z_3)}\sum_{i}t_i^{\mu\nu\rho}F_i^-(x_1,x_2,x_3),
\end{eqnarray}
where $f^{ABC}$ and $d^{ABC}$ are the anti-symmetric and symmetric structure constants of SU($N_c$). There are six tensor structures $t_i$. Their complete derivation and classification is given in appendix A of ref. \cite{Scimemi:2019gge}. Only three structures are non-vanishing for $d=4$. These are
\begin{eqnarray}\nn
t_2^{\mu\nu\rho}&=&
s^\alpha_T \epsilon^{\mu\alpha}_T g_T^{\nu\rho}
+s^\alpha_T \epsilon^{\nu\alpha}_T g_T^{\rho\mu}
+s^\alpha_T \epsilon^{\rho\alpha}_T g_T^{\mu\nu},
\\\label{def:tensor-t}
t_4^{\mu\nu\rho}&=&
-s^\alpha_T \epsilon^{\mu\alpha}_T g_T^{\nu\rho}
+2s^\alpha_T \epsilon^{\nu\alpha}_T g_T^{\rho\mu}
-s^\alpha_T \epsilon^{\rho\alpha}_T g_T^{\mu\nu},
\\\nn
t_6^{\mu\nu\rho}&=&
s^\alpha_T \epsilon^{\mu\alpha}_T g_T^{\nu\rho}
-s^\alpha_T \epsilon^{\rho\alpha}_T g_T^{\mu\nu}.
\end{eqnarray}
The other structures (i.e. $t_{3,5,7}^{\mu\nu\rho}$) parametrize evanescent operators. In general, these contributions are non-zero in the dimension regularization and should be taken into account during the renormalization procedure \cite{Dugan:1990df}. However, in the present calculation they do not contribute to the pole part, and thus decouple. For that reason these functions can be set to zero in $d=4$.

The three-gluon gluon functions are defined as \cite{Scimemi:2019gge}
\begin{eqnarray}\label{def:F24}
F_2^\pm(x_1,x_2,x_3)=-\frac{G_\pm(x_1,x_2,x_3)}{2(2-\epsilon)},
\qquad
F_4^\pm(x_1,x_2,x_3)=-\frac{Y_\pm(x_1,x_2,x_3)}{2(1-2\epsilon)}.
\label{def:F2F4}
\end{eqnarray}
The distribution $F_6$ can be expressed via $Y_\pm$
\begin{eqnarray}\label{def:F6}
F_6^\pm(x_1,x_2,x_3)=\pm\frac{Y_\pm(x_1,x_3,x_2)-Y_\pm(x_2,x_1,x_3)}{2(1-2\epsilon)}.
\end{eqnarray}
Like in the twist-two case (\ref{def:gluon-coll-d}), the $\epsilon$-dependent factors are chosen such that most of the $\epsilon$-dependence at NLO cancels.

The distributions $G_\pm$ and $Y_\pm$ satisfy the following symmetry relations
\begin{eqnarray}\nn
&&G_\pm(x_1,x_2,x_3)=
G_\pm(-x_3,-x_2,-x_1)=
\mp G_\pm(x_2,x_1,x_3)=
\mp G_\pm(x_1,x_3,x_2),
\\\label{def:sym-gluon}
&&Y_\pm(x_1,x_2,x_3)=
Y_\pm(-x_3,-x_2,-x_1)=
\mp Y_\pm(x_3,x_2,x_1),
\\\nn
&&Y_\pm(x_1,x_2,x_3)+Y_\pm(x_2,x_3,x_1)+Y_\pm(x_3,x_1,x_2)=0.
\end{eqnarray}
These relations constrain the internal structure of three-gluon distributions \cite{Scimemi:2019gge}. For a comparison of our convention with others see ref. \cite{Scimemi:2019gge}.

All twist-three distributions are functions of two variables, since the third variable is fixed by the momentum conservation condition $x_1+x_2+x_3=0$. Nevertheless, we use the the three-variable notation for its convenience since in this notation the symmetry transformations (\ref{def:sym-quark}, \ref{def:sym-gluon}) are more transparent. Also, each sector $(x_i\lessgtr 0)$ has a special interpretation in the parton picture \cite{Jaffe:1983hp}, which is harder to see in the two-variable notation. 

The set of parton distributions $\{T, \Delta T, E, H, G_\pm, Y_\pm\}$ evolves autonomously under a change of renormalization scale $\mu$ \cite{Balitsky:1987bk, Braun:2009vc},
\begin{eqnarray}\label{evol-tw3}
\mu^2\frac{d F_1(x_1,x_2,x_3;\mu)}{d\mu^2}=\sum_{F_2}\int [dy] K_{F_1\ot F_2}(x_1,x_2,x_3;y_1,y_2,y_3;a_s)F_2(y_1,y_2,y_3;\mu),
\end{eqnarray}
where $F_{1,2}\in \{T, \Delta T, E, H, G_\pm, Y_\pm\}$. Moreover, the chiral-odd distributions $E$ and $H$ do not mix with other distributions. The expressions for the evolution kernels $K_{F_1\ot F_2}$ are rather long, and not explicitly needed in the present work. For the reader's convenience we present them in position space in appendix \ref{app:evol}. The momentum space expressions are much more cumbersome \cite{Ji:2014eta}.

The set of parton distributions $\{T, \Delta T, E, H, G_\pm, Y_\pm\}$ is complete in the sense that all other twist-three distributions can be expressed in this basis (and possibly twist-two distributions). For example, the twist-three distributions $g_T$, $h_L$ and $e$ \cite{Jaffe:1996zw} can be express in terms of $\{T,\Delta T\}$, $H$ and $E$  (see e.g. \cite{Scimemi:2018mmi, Braun:2021aon, Braun:2021gvv}).

\section{Evaluation of small-$b$ expansion}
\label{sec:details}

The NLO computation presented in this work has been done using the background-field method. It is a very well developed method for the computation of perturbative corrections involving higher-twist operators. A detailed explanation of the method can be found in refs. \cite{Balitsky:1987bk, Scimemi:2019gge, Braun:2021aon, Braun:2021gvv, Vladimirov:2021hdn}. We skip the detailed description of the computation process, which can be found in ref. \cite{Scimemi:2019gge, Braun:2021aon}. In this section, we present a general discussion, and focus on particularities of the current case.

\subsection{General structure of small-$b$ expansion}

In the regime of small-$b$ the TMD operator can be expressed as a series of light-cone operators with increasing dimensions,
\begin{eqnarray}\label{OPE1}
\Phi^{[\Gamma]}(x,b)=\phi^{[\Gamma]}(x)+b^\mu \phi_{\mu}^{[\Gamma]}(x)+b^\mu b^\nu \phi_{\mu\nu}^{[\Gamma]}(x)+...~.
\end{eqnarray}
Here, the leading terms are
\begin{eqnarray}\label{phi:tw2}
\phi^{[\Gamma]}(x)&=&\frac{1}{2}\int \frac{dz}{2\pi}e^{-ixzp_+}\langle p,S|\bar q(z,n)[zn,0]\Gamma q(0)|p,S\rangle,
\\\label{phi:tw3}
\phi^{[\Gamma]}_{\mu}(x)&=&\frac{1}{2}\int \frac{dz}{2\pi}e^{-ixzp_+}\langle p,S|\bar q(z,n)[zn,-\infty n]\overleftarrow{D_\mu}[-\infty n,0]\Gamma q(0)|p,S\rangle,
\end{eqnarray}
where $D_\mu$ is the QCD covariant derivative. The series (\ref{OPE1}) is a particular application of light-cone OPE and can be written also as series of local operators \cite{Moos:2020wvd}. The matrix element (\ref{phi:tw2}) can be expresses by collinear parton distributions of twist-two, while for the matrix element (\ref{phi:tw3}) they are of twist-two and twist-three. The higher dimension matrix elements involve higher-twist distributions. 

There is no simple correspondence between the twist of TMDs and the twist of the leading contribution of its small-$b$ series. The factors $b^\mu$ in the parametrization of TMDs (\ref{def:TMDs:1:g+} -- \ref{def:TMDs:1:s+}) spoils the counting and thus the series for individual TMDs start with terms of different twist\footnote{
The coefficients in the parametrization of TMDs are not the only cause of the spoiled counting. There can be also singular contributions $\sim b^{-2}$  that appear for loop diagrams \cite{Rodini:2022wki}. However, this happens only for TMDs of higher twist.
}. So, the small-$b$ series for the TMDs $f_1$, $g_1$ and $h_1$ start with (\ref{phi:tw2}) and have leading contributions of twist-two \cite{Bacchetta:2013pqa, Echevarria:2015uaa, Gutierrez-Reyes:2017glx}. The small-$b$ series for the TMDs $f_{1T}^\perp$, $g_{1T}$, $h_{1L}^\perp$ and $h_1^\perp$ start with operators of type (\ref{phi:tw3}) and involve twist-three distributions \cite{Boer:2003cm, Kang:2011mr, Kanazawa:2015ajw, Scimemi:2018mmi}. Finally, the pretzelosity distribution $h_{1T}^\perp$ starts with $\phi_{\mu\nu}(x)$ and the leading term contains already twist-four terms \cite{Moos:2020wvd}.

The expression (\ref{OPE1}) is a tree-level expression. Accounting of quantum corrections modifies (\ref{OPE1}) by terms $\sim a_s=\alpha_s/4\pi$. These terms can be absorbed into the coefficient functions, which enter in convolution with collinear distributions. For example, the twist-two term turns into
\begin{eqnarray}\label{OPE-coef-f}
\phi_f^{[\Gamma]}(x)\to \sum_{f'} \int_x^1 \frac{dy}{y} C_{f\ot f'}(y,\ln b^2;\mu,\zeta;\mu_{\text{OPE}})\phi_{f'}^{[\Gamma]}\(\frac{x}{y},\mu_{\text{OPE}}\),
\end{eqnarray}
where indices $f$ label contributions of different parton content. The coefficient function explicitly contains the dependence on $(\mu,\zeta)$. It also contains the $\mu_{\text{OPE}}$-scale, which is the scale of OPE. The whole expression (\ref{OPE-coef-f}) is independent on $\mu_{\text{OPE}}$. Using the TMD evolution equations (\ref{TMD-evol}) and the evolution equation for collinear distributions (\ref{evol-tw2}), one can deduce the part of the coefficient function proportional to logarithms (see e.g. \cite{Echevarria:2016scs}). In what follows, we set $\mu_{\text{OPE}}=\mu$ for simplicity, such that the coefficient function depends only on $(a_s(\mu), \mathbf{L}_b, \mathbf{l}_\zeta)$. Therefore, the small-$b$ expansion for the TMDs $F\in\{f_1, g_1, h_1\}$ takes the form
\begin{eqnarray}
F_f(x,b;\mu,\zeta)=\sum_{f'} \int_x^1 \frac{dy}{y} C^{F}_{f\ot f'}(y;\mathbf{L}_b,\mathbf{l}_\zeta)f_{f'}\(\frac{x}{y},\mu\)+\mathcal{O}(b^2),
\end{eqnarray}
with $f$ being collinear distributions of twist-two. 

The expressions for twist-three have a similar general structure, but a more involved form. Generally, for $F\in \{f_{1T}^\perp, g_{1T}, h_{1L}^\perp, h_1^\perp\}$ one has
\begin{eqnarray}
F_f(x,b;\mu,\zeta)&=&\sum_{f'} \int_x^1 \frac{dy}{y} C^{F,\text{tw2}}_{f\ot f'}(y;\mathbf{L}_b,\mathbf{l}_\zeta)f_{f'}\(\frac{x}{y},\mu\)
\\\nn && +\sum_{f'} \int [dx] C^{F,\text{tw3}}_{f\ot f'}(x,x_1,x_2,x_3;\mathbf{L}_b,\mathbf{l}_\zeta)t_{f'}(x_1,x_2,x_3;\mu),
\end{eqnarray}
where $f$ and $t$ are distributions of twist-two and three, correspondingly. Note, that in the case of the Sivers and Boer-Mulders function $C^{\text{tw2}}=0$. The coefficient functions for the Sivers function are known at NLO \cite{Scimemi:2019gge}. For the other functions they are known at LO \cite{Boer:2003cm, Kang:2011mr, Kanazawa:2015ajw, Scimemi:2018mmi}, and computed here at NLO.

\subsection{Computation}

In a nutshell, the computation within the background-field method consists in following steps.
\begin{enumerate}
\item The matrix element for a TMD is presented in a functional-integral form. Then the QCD fields are split into the quantum and background modes ($q(x)=q_{\text{quan.}}(x)+q_{\text{back.}}(x)$), with corresponding momentum counting. 
\item The quantum modes are (functionally) integrated using both the perturbative expansion and the expansion in the number of background fields. The Lagrangian of the quantum-to-background fields interaction can be found in ref. \cite{Abbott:1981ke}. As result of the integration, one obtains the effective operator.
\item The effective operator is decomposed in the basis of definite-twist operators using equations of motion and algebraic manipulations.
\end{enumerate}
During this procedure one expects that the hadron is composed of the low-energy fields only, and that thus the highly-energetic quantum modes do not contribute to its wave function. Therefore, the computation is done on the level of the operator itself without any reference to the hadron state. For a detailed discussion of each step in the concrete application to TMD operators (Sivers function) we refer to \cite{Scimemi:2019gge}.

\begin{figure}[t]
\begin{center}
\includegraphics[width=0.9\textwidth]{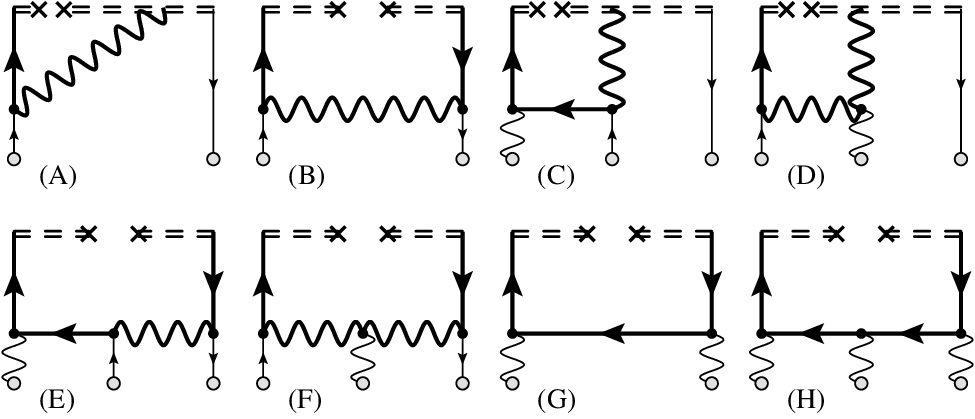}
\caption{\label{fig:diags}
Diagrams contributing to the NLO effective operator at twist-two and twist-three level. The dashed lines show the half-infinite Wilson lines. The mirror diagrams to (A, C, D, E) should be added.}
\end{center}
\end{figure}

At the twist-three level one has to compute all diagrams of mass-dimension four. They are shown in fig.\ref{fig:diags}. The diagrams with two external fields (A, B, G) have to be computed up to a single transverse-derivative contribution. These diagrams contain twist-two and twist-three parts, which can be identified using the QCD equations of motion. The diagrams with three external fields (C, D, E, F, H) contain only twist-three terms.

The diagrams have been evaluated in position space. It is the preferred representation for dealing with higher-twist operators, because the resulting expressions are much shorter in comparison to momentum space. Examples of diagram computations in this technique can be found in appendices of refs. \cite{Scimemi:2019gge, Braun:2021aon, Vladimirov:2021hdn}. The final expressions in position space are presented in appendix \ref{app:pos}. The subsequent Fourier transformation to momentum space is laborious but straightforward.

As a by-product of the computations for diagrams A and B, we obtained the NLO matching coefficients for the TMDs $f_1$, $g_1$ and $h_1$.  Our expressions coincides with well-known results  \cite{Bacchetta:2013pqa, Echevarria:2015uaa, Gutierrez-Reyes:2017glx, Buffing:2017mqm}. This served as an intermediate check of our computation.

The computation is done for the bare operators and requires renormalization. Schematically the renormalization factor has the form
\begin{eqnarray}
\Phi_{\text{renor.}}(\mu,\zeta)=
Z^{-1}_{UV}(\mu,\zeta)R^{-1}(\zeta)\Phi_{\text{bare}}
=
Z^{-1}_{UV}(\mu,\zeta)R^{-1}(\zeta)\(C_{\text{bare}}\otimes \phi_{\text{bare}}+...\),
\end{eqnarray}
where in the last equality we inserted the bare small-$b$ expansion. Here, $Z_{UV}$ is the ultraviolet renormalization factor, and $R$ is the rapidity renormalization factor. We also renormalize the collinear distribution and obtain
\begin{eqnarray}
\Phi_{\text{renor.}}(\mu,\zeta)=C_{\text{renor.}}(\mu,\zeta,\mu_{\text{OPE}})\otimes \phi_{\text{renor.}}(\mu_{\text{OPE}}),
\end{eqnarray}
where
\begin{eqnarray}
C_{\text{renor.}}(\mu,\zeta,\mu_{\text{OPE}})=Z^{-1}_{UV}(\mu,\zeta)R^{-1}(\zeta)C_{\text{bare}}\otimes Z_{\phi}(\mu_{\text{OPE}}),
\end{eqnarray}
where $Z_\phi$ is the renormalization factor for the collinear distribution $\phi$. The function $C_{\text{renor.}}$ is finite. 

To regularize divergences we use the combination of dimensional regularization and $\delta$-regularization (for rapidity divergences), which has been used in many TMD-related computations (see e.g. refs. \cite{Echevarria:2016scs, Echevarria:2012js, Buffing:2017mqm}). Collecting expressions for the LO renormalization factors \cite{Aybat:2011zv, Echevarria:2015byo}, we derive the following pocket formula for the renormalization of the NLO coefficient functions
\begin{eqnarray}\nn
C^{\text{NLO}}_{\text{renorm}}&=&\mu^{2\epsilon}e^{\epsilon\gamma_E} C_{\text{bare}}^{\text{NLO}}
+\Big[
\mu^{2\epsilon}e^{\epsilon\gamma_E}2 \(\frac{-b^2}{4}\)^\epsilon C_F \Gamma(-\epsilon)\(\mathbf{L}_b-\mathbf{l}_\zeta+2\ln\(\frac{\delta^+}{p^+}\)-\psi(-\epsilon)-\gamma_E\)
\\\label{renormalization} &&
-C_F\(\frac{2}{\epsilon^2}+\frac{3+2\mathbf{l}_\zeta}{\epsilon}\)-\frac{a_s}{\epsilon}\mathbb{H}\otimes\Big]C^{\text{LO}},
\end{eqnarray}
where the factors $\mu^{2\epsilon}e^{\epsilon \gamma_E}$ are the usual factors of the $\overline{\text{MS}}$-scheme, $\delta^+$ is the parameter of the $\delta$-regularization, $\epsilon$ is the parameter of the dimensional regularization ($d=4-2\epsilon$), and $\mathbb{H}$ is the LO evolution kernel for the corresponding collinear distribution. The cancellation of divergences in this combination is a very sensitive check of the computation.

\subsection{Treatment of $\gamma_5$}
\label{sec:gamma5}

The $\gamma^5$ matrix requires an additional treatment in dimensional regularization. In our computation we use the ``Larin+''-scheme introduced in ref. \cite{Gutierrez-Reyes:2017glx}. This is based on the four-dimensional identity
\begin{eqnarray}\label{larin+}
\gamma^+\gamma^5=\frac{i}{2!}\epsilon_T^{\mu\nu} \gamma^+\gamma_{\mu}\gamma_\nu.
\end{eqnarray}
The anti-symmetric tensor $\epsilon_T^{\mu\nu}$ is generalized to an arbitrary number of dimensions by means of the identity
\begin{eqnarray}\label{ee=gg}
\epsilon_T^{\mu_1\mu_2}\epsilon_T^{\nu_1\nu_2}=g_T^{\mu_1\nu_1}g_T^{\mu_2\nu_2}-g_T^{\mu_1\nu_2}g_T^{\mu_2\nu_1}.
\end{eqnarray}
This generalization is different from the ordinary Larin-scheme\footnote{
In the Larin scheme, one uses the identity $\gamma^+\gamma^5=i\epsilon^{+\mu\nu\rho} \gamma_{\mu}\gamma_\nu \gamma_\rho/3!$, and defines the 4-indices $\epsilon^{\mu\nu\rho\lambda}$ using the identity $\epsilon^{\mu_1\mu_2\mu_3\mu_4}\epsilon^{\nu_1\nu_2\nu_3\nu_4}=-g^{\mu_1\nu_1}g^{\mu_2\nu_2}g^{\mu_3\nu_3}g^{\mu_4\nu_4}+...$~. Therefore, the Larin-scheme treats all directions of the space-time on equal foot, whereas ``Larin+''-scheme (\ref{larin+}) specifically identifies two light-cone directions.
} \cite{Larin:1993tq}. The ``Larin+''-scheme is preferable to the Larin-scheme, because it preserves the TMD-twist of an operator \cite{Gutierrez-Reyes:2017glx, Rodini:2022wki}, and consequently, the structure of its divergences.

The generalization of the $\gamma^5$ matrix to $d$-dimensions could also involve a multiplication by scheme-dependent factor $Z_5$. However, there is no necessity to introduce such factor for the TMD operators, because their renormalization is independent on the $\Gamma$-structure (as long as it preserves the TMD-twist). The factor $Z_5$ in the ``Larin+''-scheme has been computed in ref. \cite{Gutierrez-Reyes:2017glx} demanding the equality between helicity and unpolarized coefficient functions,
\begin{eqnarray}\label{larin-condition}
Z_5\otimes C_{q\ot q}^{[\Gamma=\gamma^+\gamma^5]}=C_{q\ot q}^{[\Gamma=\gamma^+]}.
\end{eqnarray}
Unfortunately, up to now, no accurate generalization of this scheme to the twist-three case exists.

In this work, we use the following procedure, which allows us to (partially) by-pass the problems associated with the definition of $\gamma^5$. First of all, we note that the problem exists only for the worm-gear-T function $g_{1T}$. For the chiral-odd operators with $\Gamma=i\sigma^{\alpha+}\gamma^5$, the $\gamma^5$-factor is illusory since $i\sigma^{\alpha+}\gamma^5=-\epsilon_T^{\alpha\beta}\sigma_{\beta+}$. The twist-two part of the function $g_{1T}$ can be computed using the standard definition. For the twist-three part of $g_{1T}$, we distinguish quark and gluon contributions. For the pure quark contributions we use an anti-commuting $\gamma^5$ (which is equivalent to implementing condition (\ref{larin-condition})). For the gluon contributions (diagrams G and H) we compute the trace using (\ref{larin+}) and (\ref{ee=gg}). 

The result of this procedure (at NLO for the coefficient function) is equivalent to an $\overline{\text{MS}}$ twist-two computation. The deviations arrears at term suppressed by $\epsilon$ and at NNLO. It is straightforward to proof that the current scheme is equivalent at NLO to the ’t Hooft-Veltman-Breitenlohner-Maison \cite{tHooft:1972tcz, Breitenlohner:1977hr} scheme.

\subsection{Twist-decomposition of the $F_{\mu+}D_\alpha F_{\nu+}$ operator}
\label{sec:FDF}

The diagrams A, B, and G result in two-point operators of generic twist-three. Such operators must be rewritten in terms of definite-twist-2 and -3 operators, which can be accomplished by using Dirac algebra and equations of motion.

For the diagrams A and B, these operators have the form $\bar q(zn)[zn,0] \Gamma_T q(0)$ where $\Gamma \in \{\gamma^\mu, \gamma^\mu\gamma^5, \sigma^{\mu\nu}\}$ (with $\mu$ and $\nu$ being transverse indices), and $\bar q(zn)[zn,0] \Gamma_+ D_\mu q(0)$. The decomposition of such operators can be found in the literature, e.g. in refs. \cite{Balitsky:1987bk, Scimemi:2018mmi, Moos:2020wvd}. A typical relation has the form
\begin{eqnarray}\label{gT}
\langle p,S|\bar q(zn)[zn,0]\gamma^\mu \gamma^5q(0)|p,S\rangle &=& 2s_T^\mu M\int_{-1}^1 dx e^{ix \zeta }g_T(x)
\\\nn &=& 2s_T^\mu M\Big(\int_0^1 d\alpha \widehat{g_1}(\zeta)+2\zeta^2 \int_0^1 d\alpha \int_0^{\bar \alpha} d\beta \beta \widehat{S}^+(\bar \alpha \zeta,\beta \zeta,0)\Big),
\end{eqnarray}
where $\zeta=zp_+$, $\bar \alpha=1-\alpha$, and $\widehat{g_1}$ and $\widehat{S}_+$ are Fourier transformations of the corresponding collinear distributions (\ref{Fourier:tw2}, \ref{Fourier:tw3}). The first term in eqn. (\ref{gT}) gives the celebrated Wandzura-Wilczek relation \cite{Wandzura:1977qf}. 

For the diagram G the operator is $\mathbb{O}^{\mu\alpha\nu}(z)$  which comes from the expansion in $b$ of the leading-twist gluon TMD operator
\begin{eqnarray}
&&\mathbb{O}^{\mu\alpha\nu}(z)=F^{\mu+}(zn+b)[zn ,\pm\infty n]\lDer{D}^\alpha[\pm\infty n,0]F^{\nu+}(0)
\end{eqnarray}
where all indices are transverse and the sign $\pm$ depends on the process. We have not found the decomposition of his operator in the literature and, therefore, perform it here. 

To derive the decomposition, we have used the technique based on the spinor-helicity formalism developed in ref. \cite{Moos:2020wvd}. This formalism yields in a natural way the result written as Fourier transformation of the momentum space representation. The operator $\mathbb{O}^{\mu\alpha\nu}$ has twist-two and twist-three parts
\begin{eqnarray}
\mathbb{O}^{\mu\alpha\nu}(z)=
[\mathbb{O}^{\mu\alpha\nu}(z)]_{\text{tw2}}
+
[\mathbb{O}^{\mu\alpha\nu}(z)]_{\text{tw3}}.
\end{eqnarray}
For the twist-two part we found
\begin{eqnarray}
\langle p,S|\[\mathbb{O}^{\mu\alpha\nu}(z)\]_{\text{tw2}}|p,S\rangle & =& 
\frac{\e_T^{\mu\nu}s_T^\alpha M}{2(1-\e)(1-2\e)} \text{FDF}^{\text{tw2}}(z)
\label{FDF_tw2}\\
& =& \frac{\e_T^{\mu\nu}s_T^\alpha M}{2(1-\e)(1-2\e)} \int_0^1 d\alpha \int_{-\infty}^\infty dy e^{iy\alpha p^+ z}(\alpha p^+ y)^2   \Delta f_g(y),
\end{eqnarray}
where $\Delta f_g$ is the gluon-helicity distribution (\ref{def:gluon-coll-d}). The twist-three term contains three tensor structures,
\begin{eqnarray}
\langle p,S|\[\mathbb{O}^{\mu\alpha\nu}(z)\]_{\text{tw3}}|p,S\rangle & = &
t_2^{\mu\alpha\nu} M \ \text{FDF}^{\text{tw3}}_{2}(z)+
t_4^{\mu\alpha\nu} M \ \text{FDF}^{\text{tw3}}_{4}(z)+
t_6^{\mu\alpha\nu} M \ \text{FDF}^{\text{tw3}}_{6}(z),
\label{FDF_tw3} 
\end{eqnarray}
where
\begin{align*}
\text{FDF}^{\text{tw3}}_{2}(z) &= \mp ip_+^2 \pi \int_{-1}^1 dy F^+_2(-y,0,y)e^{iy p_+z},\\
\text{FDF}^{\text{tw3}}_{4}(z) &= \mp ip_+^2 \pi \int_{-1}^1 dy F^+_4(-y,0,y)e^{iyp_+z},\\
\text{FDF}^{\text{tw3}}_{6}(z) &= p_+^2 \int [dx] g^+(x_1,x_2,x_3) \int_0^1 du  \ta  \frac{3x_1+2x_3}{x_2^2}u^2 e^{-iux_1 p^+z} + \frac{x_3}{x_2^2}u^2e^{iux_3p^+z}  \tc \\
& +p_+^2 \sum_q \int[dx] 2T_q(x_1,x_2,x_3)\int_0^1 du u^2 e^{-ip_+zux_2},
\end{align*}
with $g^+=(2F_2^++F_4^++F_6^+)$. The tensors $t_i^{\mu\nu\rho}$ and functions $F_{2,4,6}$ are defined in eqns.(\ref{def:tensor-t}, \ref{def:F24}, \ref{def:F6}). The last term in $\text{FDF}^{\text{tw3}}_{6}$ is a consequence of the QCD equations of motion, and gives the singlet-quark contribution. (Note the sum over all active flavors.) The signs $\mp$ depend on the defining process, and are ``-''(``+'') for SIDIS (Drell-Yan).

\section{Results}
\label{sec:results}

In this section, we present the results for Sivers, Boer-Mulders and worm-gear TMDs in the small-$b$ regime at NLO. The expression for the Sivers function has been computed in ref. \cite{Scimemi:2019gge}. In this paper, we have re-evaluated it as cross-check and present it here for completeness. The intermediate results of our computation, which could be interesting for theoretical investigations, are presented in appendix \ref{app:pos}. 

In the formulas presented below we employ the notation for the logarithms defined in eqn.(\ref{def:logs}). The bar-variables are $\bar \alpha=1-\alpha$, $\bar y=1-y$, etc. The color factors are $C_F=(N_c^2-1)/2N_c$, $C_A=N_c$. For simplicity of presentation we use the delta-function form of the Mellin convolution
\begin{eqnarray}
\int_{-1}^1 dy \int_0^1 d\alpha \delta(x-\alpha y) f(\alpha,y)=
\left\{
\begin{array}{lc}\Ds
\int_{x}^1 \frac{dy}{y} f\(\frac{x}{y},y\), & x>0, \\\Ds
\int_{-x}^1 \frac{dy}{y} f\(\frac{-x}{y},-y\), & x<0.
\end{array}\right. 
\end{eqnarray}
The ``plus''-distribution is defined as usual
\begin{eqnarray}
(f(\alpha))_+=f(\alpha)-\delta(\bar \alpha)\int_0^1 d\beta f(\beta).
\end{eqnarray}

For all distributions the NLO expression has the following general form
\begin{eqnarray}\label{matching-general}
F(x,b;\mu,\zeta)&=&F^{(0)}(x)+ a_s\Bigg\{C_F\(-\mathbf{L}_b^2+2\mathbf{L}_b \mathbf{l}_\zeta+3\mathbf{L}_b-\frac{\pi^2}{6}\)F^{(0)}(x)
\\\nn && 
-2\mathbf{L}_b \mathbb{H}\otimes F^{(0)}(x) +F^{(1)}(x)\Bigg\}+\mathcal{O}(a_s^2,b^2),
\end{eqnarray}
where $F^{(0)}$ is the tree-level expression, $F^{(1)}(x)$ is the finite part of the coefficient function, and $\mathbb{H}\otimes F^{(0)}$ contains the evolution kernel for the corresponding distribution,
\begin{eqnarray}
\mu^2 \frac{d F^{(0)}(x)}{d \mu^2}=2a_s \mathbb{H}\otimes F^{(0)}(x).
\end{eqnarray}
The parts proportional to the logarithms follow from the evolution equations (\ref{TMD-evol}, \ref{evol-tw2}, \ref{evol-tw3}). In each case, we found agreement between our results and the known evolution equations, see appendix \ref{app:evol}.

For practical applications, it is convenient to use the so-called optimal TMDs \cite{Scimemi:2018xaf, Scimemi:2017etj}. They are defined at $\zeta=\zeta(b,\mu)$, where $\zeta(b,\mu)$ is a null-evolution curve that passes through the saddle point of $(\gamma_F,\mathcal{D})$-field \cite{Scimemi:2018xaf}. To receive the coefficient function for optimal TMDs at NLO, it is enough to set $\mathbf{l}_\zeta$ according to
\begin{eqnarray}
-\mathbf{L}_b^2+2\mathbf{L}_b\mathbf{l}_\zeta+3\mathbf{L}_b=0.
\end{eqnarray}
Note, that the remaining dependence on $\mu$ is compensated by the evolution of collinear PDF, and thus the remaining $\mu$ is the scale of OPE $\mu_{\text{OPE}}$.

\subsection{Sivers function $f_{1T}^\perp$}

The NLO expression for the Sivers function reads
\begin{eqnarray}\label{sivers:nlo}
f_{1T,q}^\perp(x,b;\mu,\zeta)&=&\pm \pi T_q(-x,0,x)\pm\pi a_s\Big\{C_F\(-\mathbf{L}_b^2+2\mathbf{L}_b \mathbf{l}_\zeta+3\mathbf{L}_b-\frac{\pi^2}{6}\)T_q(-x,0,x)
\\\nn && 
-2\mathbf{L}_b \mathbb{H}\otimes T_q(-x,0,x) +\mathbf{\delta f}_{1T}^\perp(x)\Big\}+\mathcal{O}(a_s^2,b^2).
\end{eqnarray}
The finite part is
\begin{eqnarray}\label{sivers:finite}
\mathbf{\delta f}_{1T}^\perp(x)&=&
\int_{-1}^1 dy \int_0^1 d\alpha \delta(x-\alpha y)
\Big[
\\\nn &&
\(C_F-\frac{C_A}{2}\)2\bar \alpha T_q(-y,0,y)+\frac{3 \alpha\bar \alpha}{2}\frac{G_+(-y,0,y)+G_-(-y,0,y)}{y}\Big].
\end{eqnarray}
The action of the evolution kernel on the function $T(-x,0,x)$ is
\begin{eqnarray}\label{sivers:H}
&&\mathbb{H}\otimes T_q(-x,0,x)=\int_{-1}^1 dy \int_0^1 d\alpha \delta(x-\alpha y)
\Bigg\{
\\\nn && 
\qquad \(C_F-\frac{C_A}{2}\)\Big[
\(\frac{1+\alpha^2}{1-\alpha}\)_+T_q(-y,0,y)+(2\alpha-1)_+T_q(-x,y,x-y)-\Delta T_q(-x,y,x-y)\Big]
\\\nn &&
\qquad
+\frac{C_A}{2}\Big[\(\frac{1+\alpha}{1-\alpha}\)_+T_q(-x,x-y,y)+\Delta T_q(-x,x-y,y)\Big]
\\\nn &&
\qquad
+\frac{1-2\alpha\bar \alpha}{4}\frac{G_+(-y,0,y)+Y_+(-y,0,y)+G_-(-y,0,y)+Y_-(-y,0,y)}{y}\Bigg\},
\end{eqnarray}
The choice of the sign $\pm$ is related to the process. For the case of Drell-Yan definition the "$+$" sign should be taken. For the case of SIDIS definition "$-$" sign should be taken.

In the present form, the NLO matching for the Sivers function (\ref{sivers:nlo}) has been first computed in ref. \cite{Scimemi:2019gge}.  The logarithmic part (\ref{sivers:H}) has been derived in ref. \cite{Braun:2009mi}. The quark and gluon contributions to the finite part (\ref{sivers:finite}) were derived earlier in \cite{Sun:2013hua} and \cite{Dai:2014ala}, respectively, performing fixed-order computations for the SSA cross-sections. The detailed comparison of (\ref{sivers:nlo}) with earlier work is given in ref. \cite{Scimemi:2019gge}.  In this contribution we have reproduced the results of \cite{Scimemi:2019gge} which served us as a check of our computation.

\subsection{Worm-gear-T function $g_{1T}$}

The expression for the worm-gear-T function is the most cumbersome in this work. It is convenient to split it into twist-two and twist-three contributions
\begin{eqnarray}
g_{1T,q}(x,b;\mu,\zeta)=
g_{1T,q}^{\text{tw2}}(x,b;\mu,\zeta)
+
g_{1T,q}^{\text{tw3}}(x,b;\mu,\zeta).
\end{eqnarray}

The twist-two part is convenient to present in the form
\begin{eqnarray}\label{wgt:tw2}
g_{1T,q}^{\text{tw2}}(x,b;\mu,\zeta)&=&
x\int_x^1 \frac{dy}{y} \Big[
C_{1T,q\ot q}^{\text{tw2}}\(\frac{x}{y}\)g_{1q}(y)
+
C_{1T,q\ot g}^{\text{tw2}}\(\frac{x}{y}\)\Delta f_{g}(y)\Big],
\end{eqnarray}
where
\begin{eqnarray}\nn
C_{1T,q\ot q}^{\text{tw2}}(x)&=&1+a_sC_F\[-\mathbf{L}_b^2+2\mathbf{L}_b \mathbf{l}_\zeta
-2\mathbf{L}_b\(-\bar x+2\ln\bar x-\ln x\)
-2\bar x-2\ln x-\frac{\pi^2}{6}
\]+\mathcal{O}(a_s^2),
\\
C_{1T,q\ot g}^{\text{tw2}}(x)&=&
\frac{a_s}{2}\[-2\mathbf{L}_b(2\bar x+\ln x)+2\bar x+\ln x\]+\mathcal{O}(a_s^2).
\end{eqnarray}
These expressions can be used as the Wandzura-Wilczek approximation for the worm-gear-T function. The logarithmic part of eqn. (\ref{wgt:tw2}) coincides with the one predicted by evolution equations for helicity distributions (see e.g. \cite{Moch:2014sna}).

The twist-three part is complicated. We split it into a number of terms
\begin{eqnarray}\label{wgt:tw3:main}
&&g_{1T,q}^{\text{tw3}}(x,b;\mu,\zeta)=g_{1T,q}^{(0),\text{tw3}}(x)+a_s\Big\{C_F\(-\mathbf{L}_b^2+2\mathbf{L}_b \mathbf{l}_\zeta+3\mathbf{L}_b-\frac{\pi^2}{6}\)g_{1T,q}^{(0),\text{tw3}}(x)
\\\nn && \qquad
-2\mathbf{L}_b \(\mathbb{H}_{NS}+\mathbb{H}_{G}+\sum_{q'}\mathbb{H}^{q'}_{S}\)\otimes g_{1T,q}^{\perp,(0),\text{tw3}}(x) 
+\mathbf{\delta g}_{NS}(x)
+\mathbf{\delta g}_{G}(x)
\Big\}+\mathcal{O}(a_s^2,b^2).
\end{eqnarray}
We emphasize that the singlet-quark contribution to the finite part vanishes. 

At the tree-level 
\begin{eqnarray}\label{wgt:tree}
g_{1T,q}^{(0),\text{tw3}}(x)= 2x\int [dy] \int_0^1 d\alpha 
\delta(x-\alpha y_3)\(\frac{\Delta T_q(y_{1,2,3})}{y_2^2}+\frac{T_q(y_{1,2,3})-\Delta T_q(y_{1,2,3})}{2y_2y_3}\),
\end{eqnarray}
where $(y_{i,j,k})$ is a shorthand notation for $(y_i,y_j,y_k)$. In this form the expression (\ref{wgt:tree}) has been derived in ref. \cite{Scimemi:2018mmi}. The same result (but in a different basis) has been also derived in ref. \cite{Kanazawa:2015ajw}.

The finite parts for eqn.(\ref{wgt:tw3:main}) are
\begin{eqnarray}\nn
\mathbf{\delta g}_{NS}(x)&=&
2\int[dy]\int_0^1 d\alpha \Bigg\{
\(C_F-\frac{C_A}{2}\)\(-\frac{\bar \alpha}{y_3}T+\frac{\bar \alpha(1-2\alpha)}{y_3}\Delta T\)\delta(x-\alpha y_2)
\\ &&\qquad \label{wgt:1}
+\delta(x-\alpha y_3)\Bigg[
\(-C_F\frac{\alpha \ln \alpha}{y_2}+\(C_F-\frac{C_A}{2}\)\frac{\bar \alpha y_3}{y_1 y_2}\)T
\\\nn &&\qquad
+\(C_F \frac{\alpha \ln\alpha(y_2-2y_3)-2\bar \alpha y_3}{y_2^2}+\(C_F-\frac{C_A}{2}\)\(
\frac{\bar \alpha (1-2\alpha) y_3}{y_1y_2}+\frac{2\bar \alpha^2y_3}{y_2^2}\)\)\Delta T\Bigg]\Bigg\},
\\
\mathbf{\delta g}_{G}(x)&=&
\int[dy]
\int_0^1 d\alpha \delta(x-\alpha y_3)\Bigg\{
\\\nn &&
\qquad
\alpha(\ln\alpha-2\bar \alpha)\(\frac{G_+(y_{1,2,3})-4Y_+(y_{2,3,1})}{y_2 y_3}+2\frac{Y_+(y_{2,3,1})-Y_+(y_{3,1,2})}{y_2^2}\)
\\\nn &&
\qquad
+\alpha \bar \alpha \(
8\frac{Y_+(y_{2,3,1})-Y_+(y_{3,1,2})}{y_2^2}-18\frac{Y_{+}(y_{2,3,1})}{y_2y_3}\)
\\\nn &&
\qquad
+\bar \alpha\(1-\frac{3}{8}\alpha\)\frac{-G_+(y_{1,2,3})+G_-(y_{1,2,3})+2Y_+(y_{1,2,3})+2Y_-(y_{1,2,3})}{y_1y_2}\Bigg\},
\end{eqnarray}
where we use the shortened notation $T=T_q(y_1,y_2,y_3)$, $\Delta T=\Delta T_q(y_1,y_2,y_3)$ for the quark-gluon-quark distributions.  Notice that the singlet quark contribution (summed over flavors) does not appear in the finite part. The logarithmic parts are
\begin{eqnarray}
\mathbb{H}_{NS}\otimes g_{1T,q}^{(0),\text{tw3}}(x)&=&
\int [dy]\int_0^1 d\alpha \Bigg\{
\delta(x-\alpha y_3)\Bigg[
\\\nn &&
2x C_F\Big\{\(\frac{1}{2}+\alpha-\ln\alpha+2\ln\bar \alpha\)\(\frac{T-\Delta T}{2y_2y_3}+\frac{\Delta T}{y_2^2}\)-\frac{\Delta T}{y_2^2}\Big\}
\\\nn &&
+\(C_F-\frac{C_A}{2}\)
\(\alpha \frac{(2-\alpha)T-(4-3\alpha)\Delta T}{y_2}-\bar \alpha 
\frac{T-(1-2\alpha)\Delta T}{y_1}\)
\\\nn &&
+\frac{C_A}{2}\Big\{\(\frac{\alpha \bar \alpha-2}{y_2}-\frac{1}{x+y_1}\)\(T-\Delta T-2y_3 \frac{\Delta T}{y_2}\)-2(1-2\alpha)y_3\frac{\Delta T}{y_2^2}\Big\}\Bigg]
\\\nn &&
+\delta(x-\alpha y_2)\(C_F-\frac{C_A}{2}\)\(-\alpha +\bar \alpha^2\frac{y_2}{y_3}\)\frac{T+(1-2\alpha)\Delta T}{x+y_1}
\\\nn &&
+\delta(x-y_2-\alpha y_3)\(C_F-\frac{C_A}{2}\)\frac{1}{y_2}\[
T+\(1+2\frac{\alpha y_3}{y_2}\)\Delta T\]
\Bigg\},
\end{eqnarray}
\begin{eqnarray}
\mathbb{H}_{G}\otimes g_{1T,q}^{(0),\text{tw3}}(x)&=&
-\int[dy]
\int_0^1 d\alpha \delta(x-\alpha y_3)\Bigg\{\frac{\alpha\bar \alpha}{2} \frac{Y_+(y_{2,3,1})-Y_+(y_{3,1,2})}{y_2 y_3}
\\\nn &&
\qquad
+\alpha(2\bar \alpha+\ln\alpha)\(\frac{G_+(y_{1,2,3})-2Y_+(y_{2,3,1})}{y_2 y_3}+\frac{Y_+(y_{2,3,1})-Y_+(y_{3,1,2})}{y_2^2}\)
\\\nn &&
\qquad
+\frac{\bar \alpha}{4}\frac{G_+(y_{1,2,3})-G_-(y_{1,2,3})}{y_1y_2}
-\frac{\bar \alpha (1-3\alpha)}{2}\frac{Y_+(y_{3,1,2})-Y_-(y_{3,1,2})}{y_1y_2}
\\\nn &&
\qquad
+\frac{\bar \alpha (1-2\alpha)}{2}\frac{Y_+(y_{2,3,1})-Y_+(y_{3,1,2})-Y_-(y_{2,3,1})+Y_-(y_{3,1,2})}{y_2^2}\Bigg\},
\\\label{wgt:2}
\mathbb{H}^{q'}_{S}\otimes g_{1T,q}^{\perp,(0),\text{tw3}}(x) &=&
2\int[dy]\int_0^1 d\alpha \delta(x-\alpha y_2)(\alpha\bar \alpha +\alpha\ln \alpha)\frac{T_{q'}(y_1,y_2,y_3)}{y_2},
\end{eqnarray}
where we use the shortened notation $T=T_q(y_1,y_2,y_3)$, $\Delta T=\Delta T_q(y_1,y_2,y_3)$ for the quark-gluon-quark distributions, and $(y_{i,j,k})=(y_i,y_j,y_k)$ for three-gluon distributions. To simplify these expressions we have used the symmetry relations (\ref{def:sym-quark}) and (\ref{def:sym-gluon}). 

The logarithmic part coincides with the prediction given by the renormalization group equation \cite{Braun:2009mi, Braun:2009vc} (see appendix \ref{app:evol}). It provides a strong check of our computation. The comparison has been made in position space (see appendix \ref{app:pos}). The integrands of eqns. (\ref{wgt:1} -- \ref{wgt:2}) are finite for $y_i\to0$. Also, we observed the cancelation of various undesirable terms such as $\ln^2\alpha$ and $\ln \bar\alpha/\alpha$ that appear in the individual diagrams. Altogether, these observations provide extra confidence in the result.

\subsection{Boer-Mulders function $h_{1}^\perp$}
The Boer-Mulders function is in many aspects similar to the Sivers function, which is the consequence of their T-oddness. We have
\begin{eqnarray}
h_{1,q}^{\perp}(x,b;\mu,\zeta)&=&\mp \pi E_q(-x,0,x)\mp \pi a_s\Big\{C_F\(-\mathbf{L}_b^2+2\mathbf{L}_b \mathbf{l}_\zeta+3\mathbf{L}_b-\frac{\pi^2}{6}\)E_q(-x,0,x)
\\\nn && 
-2\mathbf{L}_b \mathbb{H}\otimes E_q(-x,0,x)  \Big\}+\mathcal{O}(a_s^2,b^2).
\label{h1Perp_tempLabel}
\end{eqnarray}
where the $\mp$ identifies the process under consideration. For DY (SIDIS) the upper (lower) sign should be taken. For the Boer-Mulders function, we have found that the finite part (besides the $\pi^2/6$ contribution), exactly vanishes, i.e.:
\begin{eqnarray}\label{bm:finite}
\mathbf{\delta h}_{1,f}^{\perp}(x) &=& 0
\end{eqnarray}
For the evolution kernel, we have 
\begin{eqnarray}
&& \mathbb{H}\otimes E_q(-x,0,x) =-\frac{C_F}{2}E_q(-x,0,x)+\int_0^1 d\alpha\int dy \delta(x-\alpha y) \Big\{
\\\nn &&\qquad
2\(C_F-\frac{C_A}{2}\)
\Big[\(\frac{\alpha}{1-\alpha}\)_+E_q(-y,0,y)
-\bar \alpha E_q(-x,y,x-y)\Big]
+C_A\frac{E_q(-x,x-y,y)}{(1-\alpha)_+}\Big\}.
\end{eqnarray}
In general the expression for the Boer-Mulders function has the simplest form among all TMD distributions that match twist-three operators. The expression for the evolution kernel agrees with the general kernel for the twist-three functions \cite{Braun:2009vc, Braun:2021gvv}, see also appendix \ref{app:evol}.

%

\subsection{Worm-gear-L function $h_{1L}^\perp$}

It is convenient to split the expression for the worm-gear-T function into twist-two and twist-three contributions
\begin{eqnarray}
h_{1L,q}^\perp(x,b;\mu,\zeta)=
h_{1L,q}^{\perp,\text{tw2}}(x,b;\mu,\zeta)
+
h_{1L,q}^{\perp,\text{tw3}}(x,b;\mu,\zeta).
\end{eqnarray}

The twist-two part can be written in the form
\begin{eqnarray}\label{wgl:tw2}
h_{1L,q}^{\perp,\text{tw2}}(x,b;\mu,\zeta)&=&
-x^2\int_x^1 \frac{dy}{y^2} 
C_{1L,q\ot q}^{\perp,\text{tw2}}\(\frac{x}{y}\)h_{1}(y),
\end{eqnarray}
where
\begin{eqnarray}\nn
C_{1L,q\ot q}^{\perp,\text{tw2}}(x)&=&1+a_sC_F\[-\mathbf{L}_b^2+2\mathbf{L}_b \mathbf{l}_\zeta
+4\mathbf{L}_b\(\ln x-\ln\bar  x\)-\frac{\pi^2}{6}
\]+\mathcal{O}(a_s^2),
\end{eqnarray}
These expressions can be used as the Wandzura-Wilczek-like approximation for the worm-gear-L function. The logarithmic part of eqn. (\ref{wgl:tw2}) coincides with the one predicted by evolution equations for transversity distributions (see e.g. \cite{Vogelsang:1997ak}). The finite part contains only the trivial contribution $\pi^2/6$. The non-trivial part vanishes (see the diagram $\mathbf{B}$ in sec.\ref{app:diagrams:wgh}).

The twist-three part is
\begin{eqnarray}\label{wgl:tw3:main}
h_{1L,q}^{\perp,\text{tw3}}(x,b;\mu,\zeta)&=&h_{1L,q}^{\perp,(0),\text{tw3}}(x)+a_s\Big\{C_F\(-\mathbf{L}_b^2+2\mathbf{L}_b \mathbf{l}_\zeta+3\mathbf{L}_b-\frac{\pi^2}{6}\)h_{1L,q}^{\perp,(0),\text{tw3}}(x)
\\\nn && 
-2\mathbf{L}_b \mathbb{H} \otimes h_{1L,q}^{\perp,(0),\text{tw3}}(x) 
+\mathbf{\delta h}(x)
\Big\}+\mathcal{O}(a_s^2,b^2).
\end{eqnarray}
At tree-level it is
\begin{eqnarray}
h_{1L,q}^{\perp,(0),\text{tw3}}(x)=  -2x \int_0^1 d\alpha \int[dy] \alpha \delta(x-\alpha y_3)  H_q(y_1,y_2,y_3) \frac{y_3-y_2}{y_2^2y_3}.
\end{eqnarray}
This expression has been derived in refs. \cite{Scimemi:2018mmi, Kanazawa:2015ajw}. Note that the integral is finite for $y_2\to0$, since $H(-y,0,y)=0$.

The finite and logarithmic parts of the twist-three expression are
\begin{eqnarray}
&&\mathbf{\delta h}(x)=-4\int [dy]H_q(y_1,y_2,y_3)
\Bigg\{
\\\nn &&\qquad
\int_0^1 d\alpha \Big[
\(C_F-\frac{C_A}{2}\)\alpha \bar \alpha \(\frac{\delta(x-\alpha y_2)}{y_3}-\frac{\delta(x-\alpha y_3)}{y_1}\)
+\frac{C_A}{2} \bar \alpha (\alpha y_2+\bar \alpha y_3)\frac{\delta(x-\alpha y_3)}{y_2^2}\Big]
\\\nn &&\qquad
+\int_0^1 d\alpha \int_0^1 d\beta \frac{\alpha}{x+y_1}\(\frac{C_A}{2}\delta(x+\alpha y_1+\alpha \beta y_2)-\(C_F-\frac{C_A}{2}\)\delta(x+\alpha y_1+\alpha \beta y_3)\)\Bigg\}.
\\
&& \mathbb{H} \otimes h_{1L,q}^{\perp,(0),\text{tw3}}(x) =-2\int [dy]H_q(y_1,y_2,y_3)
\Bigg\{
\\\nn &&\qquad
\int_0^1 d\alpha C_F \alpha x \(\frac{3}{2}+2 \ln \bar \alpha-2\ln \alpha\)\frac{y_3-y_2}{y_2^2 y_3} \delta(x-\alpha y_3)
\\\nn &&\qquad
+\int_0^1 d\alpha \int_0^1 d\beta \frac{\alpha (y_2-x)}{y_2(x+y_1)}\(\frac{C_A}{2}\delta(x+\alpha y_1+\alpha \beta y_2)-\(C_F-\frac{C_A}{2}\)\delta(x+\alpha y_1+\alpha \beta y_3)\)\Bigg\}.
\end{eqnarray}
The double-integrals in the last lines of these equations can be integrated over one of the variables, but the resulting expressions have a complicated form.

\section{Conclusion}

We have computed the leading small-$b$ asymptotics for Sivers ($f_{1T}^\perp$), Boer-Mulders ($h_{1}^\perp$) and worm-gear functions ($g_{1T}$ and $h_{1L}^\perp$) at NLO in perturbation theory. These functions are expressed in terms of twist-two and twist-three collinear distributions. The computation is performed using the well-established background-field method, which was also used for similar computations in refs. \cite{Scimemi:2019gge, Braun:2021aon, Braun:2021gvv}. The result is presented both in position (appendix \ref{app:pos}) and momentum-fraction (section \ref{sec:results}) space. The logarithmic parts of the obtained expressions agree with the predictions of the renormalization group equations. The result for the Sivers function coincides with the one computed in ref. \cite{Scimemi:2019gge}.

With the results of this work the knowledge of small-$b$ expressions for TMDs of leading twist is complete at NLO (or even higher, see refs. \cite{Luo:2020epw, Gutierrez-Reyes:2018iod}). The only distribution for which this is still missing is pretzelosity that has leading twist-four contributions at small-$b$ \cite{Moos:2020wvd}. In the transverse momentum space the computed expressions corresponds to the large momentum asymptotic of TMDs.

The perturbative expansions for the Sivers and Boer-Mulders functions on one side and the worm-gear functions on the other side are drastically different, which is a consequence of the T-parity properties of these functions. So, the Sivers and Boer-Mulders at LO have the Qiu-Sterman form of quark-anti-quark correlators with a null-momentum gluon field \cite{Qiu:1991pp} $T(-x,0,x)$ and $E(-x,0,x)$. The NLO expressions for these distributions contain only twist-three distributions and are relatively simple (in particular, the finite part of the Boer-Mulders function is trivial (\ref{bm:finite})). The global sign of the small-$b$ expression depends on the orientation of the gauge link.

In contrast, the worm-gear functions have involved forms. Already at LO, they are expressed by convolution integrals of twist-two and twist-three distributions, which lead to bulky NLO expressions. The expression for the worm-gear-T distribution is especially cumbersome, since it contains mixtures with a three-gluon correlator and a singlet-quark contribution. Unfortunately, we have not found any significant simplifications for these distributions. At the moment, the most practically important result for worm-gear functions is the part proportional to twist-two distributions, because it can be used as an approximation for these functions (Wandzura-Wilczek-like approximation). The goodness of such an approximation is difficult to establish at the moment. It remains, however, a useful one given the currently available data.

The derived NLO expressions are important for the phenomenology of TMDs and twist-three distributions. They provide the leading logarithmic terms, and thus allow to properly include QCD evolution effects in the data analysis. This will be definitely important for the next-generation of high-precision polarized experiments such as EIC \cite{AbdulKhalek:2021gbh}. 

\acknowledgments
We thank Vladimir Braun for discussions. A.V. is funded by the \textit{Atracci\'on de Talento Investigador} program of the Comunidad de Madrid (Spain) No. 2020-T1/TIC-20204. This work was partially supported by DFG FOR 2926 ``Next Generation pQCD for  Hadron  Structure:  Preparing  for  the  EIC'',  project number 430824754.

\appendix

\section{Evolution equations for twist-three collinear distributions}
\label{app:evol}

In this appendix, we collect the expressions for the LO evolution kernels of twist-three distributions $\widehat{F}$. The expressions are given in position space where they are more compact and which we used for the checks of our computations. We define
\begin{eqnarray}\label{Fourier:tw3}
\widehat{F}(\zeta_1,\zeta_2,\zeta_3)=\int [dx] e^{-i(\zeta_1x_1+\zeta_2x_2+\zeta_3x_3)}F(x_1,x_2,x_3),
\end{eqnarray}
for $F\in\{S^\pm, T,\Delta T, H, E, F_2, F_4, F_6\}$.

The evolution equations in position space have the form
\begin{eqnarray}
\mu^2 \frac{d}{d\mu^2}\widehat{F}(\zeta_1,\zeta_2,\zeta_3)=2a_s [\mathbb{H}\otimes \widehat{F}](\zeta_1,\zeta_2,\zeta_3),
\end{eqnarray}
where $\mathbb{H}$ is an integral operator. The derivation and original expressions for the kernels can be found in refs. \cite{Balitsky:1987bk,Braun:2009vc}.  The momentum space expressions are much longer. They can be found (in parts) in refs. \cite{Ji:2014eta, Braun:2009mi, Vladimirov:2021hdn}. 

The evolution kernel for the quark-gluon-quark chiral-even operators has three flavor contributions
\begin{eqnarray}
[\mathbb{H}\otimes \widehat{F}_q]
=
[\mathbb{H}_{NS}\otimes \widehat{F}_q]
+
[\mathbb{H}_{G}\otimes \widehat{F}_q]
+
\sum_{q'} [\mathbb{H}^{q'}_{S}\otimes \widehat{F}_{q}],
\end{eqnarray}
where $q$ labels the flavor of the quark field, and we omit the arguments $(z_1,z_2,z_3)$ in each term. The non-singlet part for the function $\widehat{S}^+$ reads
\begin{eqnarray}
&&[\mathbb{H}_{NS}\otimes \widehat{S}^+](\zeta_1,\zeta_2,\zeta_3)
\\\nn &&\quad= 
\frac{C_A}{2}\int_0^1 d\alpha \Big[
\frac{\bar\alpha \widehat{S}^+(\zeta_{12}^\alpha,\zeta_2,\zeta_3)+\bar\alpha \widehat{S}^+(\zeta_1,\zeta_2,\zeta_{32}^\alpha) +\bar\alpha^2 \widehat{S}^+(\zeta_1,\zeta_{21}^\alpha,\zeta_3)+\bar\alpha^2 \widehat{S}^+(\zeta_1,\zeta_{23}^\alpha,\zeta_3)}{(\alpha)_+}
\\\nn &&\qquad +2 \int_0^{\bar \alpha}d\beta \bar \beta \widehat{S}^+(\zeta_{12}^\alpha,\zeta_{21}^\beta,\zeta_3)\Big] 
\\\nn &&\qquad+\(C_F-\frac{C_A}{2}\)\int_0^1 d\alpha\Big[
\frac{\bar\alpha \widehat{S}^+(\zeta_{13}^\alpha,\zeta_2,\zeta_3)+\bar \alpha \widehat{S}^+(\zeta_1,\zeta_2,\zeta_{31}^\alpha)}{(\alpha)_+}-\bar \alpha \widehat{S}^+(\zeta_1,\zeta_{32}^\alpha,\zeta_3)
\\\nn &&\qquad 
+\int_0^{\bar \alpha}d\beta \widehat{S}^+(\zeta_{13}^\alpha,\zeta_2,\zeta_{31}^\beta)
-2\int_{\bar \alpha}^1 d\beta \bar \beta \widehat{S}^+(\zeta_{12}^\alpha,\zeta_{21}^\beta,\zeta_3)\Big]
+\frac{3}{2}C_F \widehat{S}^+(\zeta_1,\zeta_2,\zeta_3),
\end{eqnarray}
where 
\begin{eqnarray}\nn
\bar \alpha=1-\alpha,\qquad \zeta_{ij}=\zeta_i \bar \alpha+\zeta_j \alpha.
\end{eqnarray}
The gluon mixing (also for the $\widehat{S}^+$ function) is
\begin{eqnarray}
&&[\mathbb{H}_{G}\otimes \widehat{S}^+](z_1,z_2,z_3)
\\\nn &&\quad=-i(\zeta_1-\zeta_3)\Big[
\int_0^1 d\alpha \int_0^{\bar \alpha}d\beta (1-\alpha-\beta+2\alpha\beta)(g^++g^-)
+
\int_0^1 d\alpha \int_{\bar \alpha}^1d\beta \bar \alpha \bar \beta (-g^++g^-)\Big],
\end{eqnarray}
where
\begin{eqnarray}
g^\pm=2 \widehat{F}_2^\pm(\zeta_{13}^\alpha,\zeta_2,\zeta_{31}^\beta)
+\widehat{F}_4^\pm(\zeta_{13}^\alpha,\zeta_2,\zeta_{31}^\beta)
+\widehat{F}_6^\pm(\zeta_{13}^\alpha,\zeta_2,\zeta_{31}^\beta),
\end{eqnarray}
with $\widehat{F}_{1,2,3}$ being defined in eqn.(\ref{def:F2F4}, \ref{def:F6}). Finally, the mixture with the quark-gluon-quark operators (of all active flavors including the original one) is
\begin{eqnarray}
&&[\mathbb{H}_{S}^{q}\otimes \widehat{S}^+](\zeta_1,\zeta_2,\zeta_3)
=\int_0^1 d\alpha \,\alpha\bar \alpha \widehat{S}^+(\zeta_{13}^\alpha,\zeta_2,\zeta_{13}^\alpha).
\end{eqnarray}
This contribution appears via the QCD equation of motion in the diagrams with external ``bad'' components of gluon fields (see e.g. \cite{Ji:2001bm}).

The evolution kernel of the chiral-odd functions  is
\begin{eqnarray}
&&[\mathbb{H}\otimes \widehat{F}](\zeta_1,\zeta_2,\zeta_3)
\\\nn &&\quad= 
\frac{C_A}{2}\int_0^1 d\alpha \Big[
\frac{\bar\alpha \widehat{F}(\zeta_{12}^\alpha,\zeta_2,\zeta_3)+\bar\alpha \widehat{F}(\zeta_1,\zeta_2,\zeta_{32}^\alpha) +\bar\alpha^2 \widehat{F}(\zeta_1,\zeta_{21}^\alpha,\zeta_3)+\bar\alpha^2 \widehat{F}(\zeta_1,\zeta_{23}^\alpha,\zeta_3)}{(\alpha)_+}
\\\nn &&\qquad +2 \int_0^{\bar \alpha}d\beta \bar \beta \(
\widehat{F}(\zeta_{12}^\alpha,\zeta_{21}^\beta,\zeta_3)
+\widehat{F}(\zeta_1,\zeta_{23}^\beta,\zeta_{32}^\alpha)\)\Big] 
\\\nn &&\qquad+\(C_F-\frac{C_A}{2}\)\int_0^1 d\alpha\Big[
\frac{\bar\alpha \widehat{F}(\zeta_{13}^\alpha,\zeta_2,\zeta_3)+\bar \alpha \widehat{F}(\zeta_1,\zeta_2,\zeta_{31}^\alpha)}{(\alpha)_+}
\\\nn &&\qquad -2\int_{\bar \alpha}^1 d\beta \bar \beta \(
\widehat{F}(\zeta_{12}^\alpha,\zeta_{21}^\beta,\zeta_3)
+\widehat{F}(\zeta_1,\zeta_{23}^\beta,\zeta_{31}^\alpha)\)\Big]
+\frac{3}{2}C_F \widehat{F}(\zeta_1,\zeta_2,\zeta_3),
\end{eqnarray}
where $\widehat{F}$ stands for  $\widehat{H}$ or $\widehat{E}$. Note, that the equation can be simplified for each case using (anti)symmetry of the functions $\widehat{E}(\widehat{H})$.

\section{Intermediate expressions in position space}
\label{app:pos}
\newcommand{\z}{\zeta}
In this appendix we provide the full set of expressions in position space obtained by evaluating the diagrams
with the background field method. 
For the twist-two distributions, $F\in \{f_1, g_1, h_1, f_g, \Delta f_g\}$ we define
\begin{eqnarray}\label{Fourier:tw2}
\widehat{F}(\zeta)=\int_{-1}^1 dx e^{ix \zeta}F(x).
\end{eqnarray}
For the twist-three distributions $F\in\{S^\pm, T,\Delta T, H, E, G, Y\}$ we define
\begin{eqnarray}\label{Fourier:tw3}
\widehat{F}(\zeta_1,\zeta_2,\zeta_3)=\int [dx] e^{-i(\zeta_1 x_1+\zeta_2x_2+\zeta_3x_3)}F(x_1,x_2,x_3),
\end{eqnarray}
being $[dx] = dx_1 dx_2dx_3 \delta(x_1+x_2+x_3)$.
In position space the collinear distributions satisfy translation invariance
\begin{eqnarray}
\widehat{F}(\zeta_1+\tau,\zeta_2+\tau,\zeta_3+\tau)=\widehat{F}(\zeta_1,\zeta_2,\zeta_3).
\end{eqnarray}
In the following formulas, we have used this relation together with the symmetry relations to simplify expressions.

In the following we use the notation $[d\alpha d\beta d\gamma]$ to denote the integral over the simplex of Feynman variables, i.e.
\begin{equation}
\int [d\alpha d\beta d\gamma] = \int_0^1 d\alpha \int_0^1 d\beta \int_0^1d\gamma \delta(1-\alpha-\beta-\gamma)
\end{equation}
We present the results for both SIDIS- and DY-like TMDs. For this reason, it is convenient to introduce $L$ as
\begin{equation}
L = \begin{cases}
+\infty & \text{SIDIS-like process}, \\
-\infty & \text{DY-like process}.
\end{cases}
\end{equation}
For all diagrams we show the contribution to a particular TMD. For example, for $g_{1T}$ we extract the coefficient of $(b\cdot s_T)$, and divide it by $iM$.

\subsection{Worm-gear-T function $g_{1T}$}
Diagrams A and B are most conveniently written in terms of the tree level expressions for the matching of the worm-gear function $g_{1T}$ and for the function $g_T$. In position space they are:
\begin{equation}
\begin{split}
\widehat{g}_{1T}^{tree}(\zeta) &= \frac{1}{i\zeta}\( \widehat{g}_1(\zeta) + \int_0^1d\beta \widehat{g}_1(\beta \zeta)\) + i\( \int_L^0 d\tau- \int_0^{-L}d\tau\)  \widehat{S}^+(\zeta,\tau,0) \\
&+ i\zeta\int [d\alpha d\beta d\gamma ]\(2\beta \widehat{S}^+(\bar{\alpha}\z,\beta \z )+2\widehat{S}^+(\zeta,\beta\zeta,0)\),
  \\
\widehat{g}_T(\zeta) &= \int_0^1 d\alpha \widehat{g}_1(\alpha\zeta) + 2\zeta^2 \int [d\alpha d\beta d\gamma] \beta \widehat{S}^+(\bar{\alpha}\zeta,\beta\zeta,0).
\end{split}
\end{equation}
We use the distributions $S^+, T, \Delta T$ to present the results. These are linked by the following relations
\begin{eqnarray}
\widehat{T}(\zeta_1,\zeta_2,\zeta_3) &=& \widehat{T}(-\zeta_3,-\zeta_2,-\zeta_1), \quad  \widehat{\Delta T}(\zeta_1,\zeta_2,\zeta_3) = -\widehat{\Delta T}(-\zeta_3,-\zeta_2,-\zeta_1), \\
2 \widehat{S}^\pm(\zeta_1,\zeta_2,\zeta_3) &=& -\widehat{T}(\zeta_1,\zeta_2,\zeta_3) \pm \widehat{\Delta T}(\zeta_1,\zeta_2,\zeta_3), \quad \widehat{S}^+(\zeta_1,\zeta_2,\zeta_3) = \widehat{S}^-(-\zeta_3,-\zeta_2,-\zeta_1)
\end{eqnarray}

On a diagram by diagram basis, we have for $g_{1T}$:
\begin{eqnarray}
\bm{A+A^*} &=& 2a_s C_F B^\e \Gamma(-\e) \int_0^1 d\alpha \(\frac{2\alpha}{\bar{\alpha}}\)_+ \alpha \widehat{g}_{1T}^{tree}(\alpha\zeta) - 2\delta(\bar{\alpha})(1+\lambda_\delta)\widehat{g}_{1T}^{tree}(\alpha\zeta),\\
\bm{B} &=& 2a_s(1-\e)\Gamma(-\e)C_F B^\e \int_0^1 d\alpha \  2\alpha\bar{\alpha}\hat{g}^{tree}_{1T}(\alpha \zeta)+ (1-2\alpha) \frac{\widehat{g}_T(\alpha \zeta)}{i\zeta},\\
\bm{C+C^*} &=&  i a_s\( C_F-\frac{C_A}{2}\) \Gamma(-\e) B^\e  \int [d\alpha d\beta d\gamma] \left[\int_L^0 d\tau- \int_0^{-L} d\tau\right] \\
&&\times \Bigg\{ 2\frac{\beta}{\bar{\beta}} \widehat{\Delta T}\(\beta \z; \tau -\alpha \z;0\) \nn\\
&& - \(2\frac{\alpha}{\bar{\beta}}-1\) \widehat{\Delta T}\( \zeta_{\tau \zeta}^\beta; \alpha (\tau-\z); 0\)  + \widehat{T}\( \zeta_{\tau \zeta}^\beta; \alpha (\tau-\z); 0\) 
\Bigg\}, \nn \\
\bm{D+D^*} &=& -i a_s \frac{C_A}{2}\Gamma(-\e)B^\e\(\int_L^0d\tau-\int_0^{-L}d\tau\)  \int [d\alpha d\beta d\gamma] \\
&& \times \Bigg[ \( 1+2\frac{\alpha}{\beta}\)\widehat{\Delta T}(\zeta_{\zeta \tau}^\beta,\alpha(\z-\tau),0)+ \widehat{T}(\zeta_{\zeta \tau}^\beta,\alpha(\z-\tau),0) \nn\\
&& +2 \frac{\bar{\beta}}{\beta}  \widehat{\Delta T}(\bar{\beta}\z, \tau + \alpha \z ,0) \Bigg], \nn\\
\bm{E+E^*} &=& -ia_s \Gamma(-\e)B^\e \( C_F-\frac{C_A}{2}\)  \int [d\alpha d\beta d\gamma] \( \int_L^0 d\tau - \int _0^{-L}d\tau\) \\
&& \times \Bigg[ (1-\e)(1-4\gamma)
\widehat{\Delta T}\( \alpha \z;\tau-\beta \z;0\)  - (1+\e)
\widehat{T}\( \alpha \z;\tau- \beta \z;0\) 
\Bigg]
\Bigg],\nn\\
\bm{F} &=& -2ia_s  \frac{C_A}{2}\Gamma(-\e)(1-\e)B^\e\int [d\alpha d\beta d\gamma] \beta \left(\int_L^0d\tau -\int_0^{-L}d\tau\right)\widehat{\Delta T}(\bar{\beta} \z,\tau+\alpha \z,0),
\\
\bm{G} &=&  i a_s\Gamma(-\e) B^\e \int_0^1 d\alpha \Bigg[\frac{\alpha}{2}\(2\alpha-1 - 2\bar{\alpha} \frac{\e}{1-\e}\)
\(\int_L^0d\tau - \int_0^{-L}d\tau\)   \text{FDF}^{\text{tw2}}(\tau+\alpha \z)  \notag \\
&&+(1-2\e) \alpha(2\alpha-1-2\e\bar{\alpha})   \ta\int_L^0 d\tau - \int_0^{-L} d\tau\tc\text{FDF}^{\text{tw3}}_6(\tau+\alpha \zeta)
\\
&& -i(1-2\e)^2\frac{\bar{\alpha}\alpha}{2}\ta \int_L^0d\tau - \int_0^{-L}d\tau\tc\ta \int_L^0d\sigma - \int_0^{-L}d\sigma\tc F_6(\sigma+\alpha \zeta,\tau,0)
\notag \\
&&  -2(1-2\e) \(\int_L^0 d\tau-\int_0^{-L}d\tau\) \alpha \bar{\alpha} \widehat{T}(0,\tau+\alpha \z,0)\Bigg]
\nn. 
\end{eqnarray}

For diagram H, we present the result using light-cone gauge for the background fields, which allows us to write
\begin{equation}
A^{\mu}(z) = -\int_L^0 d\tau F^{\mu+}(\tau + z).
\end{equation}
For more details on this relation, we refer to Ref.~\cite{Vladimirov:2021hdn}. Also, to present the result in a compact form, we define $\partial_{1,2,3}$ as derivatives acting only on $A^\mu$, $A^\nu$ and $A^\sigma$, respectively.
We obtain:
\begin{align}
\bm{H} &= \frac{g a_s \Gamma(-\e)B^\e}{4i(b\cdot s_T)M} \int [d\alpha d\beta d\gamma d\rho] (d^{ABC}+if^{ABC})\braket{P|A^{\mu}_A(\beta\z) A^\nu_B((\beta+\gamma) \z) A^\sigma_C(\bar{\alpha}\z)|P}\\
& + i b^{\sigma } (\partial_1^+-\partial_3^+) \e_T^{\mu \nu } (\gamma  \z \partial_1^+ - \rho\z\partial_3^+   -1)
\nn \\
& 
+i b^{\nu } \e_T^{\mu \sigma } \left(\gamma  (2 \gamma -1) \z(\partial_1^+)^2 
 + \rho  (2 \rho -1)\z (\partial_3^+)^2 
-8 \gamma\partial_1^+ 
+8\rho \partial_3^+
-\partial_1^+\partial_3^+ \z  (4 \gamma  \rho -\gamma-\rho)+4\partial_1^++\partial_2^+\right)
\nn \\
& -i b^{\mu } \e_T^{\nu \sigma } \left(
\gamma  (2 \gamma -1)  \z(\partial_1^+)^2 
+\rho (2\rho-1) \z(\partial_3^+)^2 
-8 \gamma  \partial_1^+
+8 \rho  \partial_3^+  
-\partial_1^+ \partial_3^+ \z  (4 \gamma  \rho -\gamma-\rho)-4\partial_3^+- \partial_2^+\right)
\nn \\
& +i \e_T^{\mu \rho}b_\rho \left( g^{\nu \sigma } \(
\rho  \z \partial_2^+\partial_3^+
-\gamma  \z \partial_1^+ \partial_2^+ 
+\partial_2^+ -2 \partial_3^+\)
+4 \epsilon \partial_1^+   \frac{b^{\nu } b^{\sigma}}{b^2}\right)
\nn \\
& +i \e_T^{\nu \rho}b_\rho \left( g^{\mu \sigma } \(
\rho  \z\partial_2^+\partial_3^+
-\gamma  \z \partial_1^+  \partial_2^+
+\partial_2^+-2\partial_1^+
\) 
+4 \epsilon  \partial_3^+  \frac{b^{\mu } b^{\sigma }}{b^2}\right)
\nn \\
& +i \e_T^{\sigma \rho}b_\rho \left( g^{\mu \nu } \(\gamma \z \partial_1^+\partial_2^+  -  \rho  \z \partial_2^+\partial_3^+ -\partial_2^+\)+4 \epsilon \partial_2^+ \frac{b^{\mu } b^{\nu }}{b^2}\right).
\end{align}
The factor $i(b\cdot s_T)M$ comes from the definition of $g_{1T}(\zeta,b)$. Expanding the result for diagram $\bm{H}$ and writing it in terms of distributions is most conveniently done using directly the momentum space representation.

\subsection{Boer-Mulders function $h_1^\perp$}
The Boer-Mulders function is similar to the Sivers function. It has only a twist-3 contribution. Specifically we have 
\begin{equation}
\widehat{h}_{1}^{\perp,tree}(\zeta) = -\frac{1}{2}\int_L^{-L} d\tau \widehat{E}(\zeta,\tau,0).
\end{equation}
The function $E(\zeta_1,\zeta_2,\zeta_3)$ obeys the symmetry relation
\begin{equation}
\widehat{E}(\zeta_1,\zeta_2,\zeta_3) = \widehat{E}(-\zeta_3,-\zeta_2,-\zeta_1)
\end{equation}
In this expression, it is trivial to see that, passing from SIDIS-like processes to DY-like processes, the function changes sign.

For the individual diagrams, we have
\begin{eqnarray}
\bm{A+A^*} && = 2a_s C_F B^\e \Gamma(-\e) \int_0^1 d\alpha \(\frac{2\alpha}{\bar{\alpha}}\)_+ \alpha \widehat{h}_{1}^{\perp,tree}(\alpha\zeta) - 2\delta(\bar{\alpha})(1+\lambda_\delta)\widehat{h}_{1}^{\perp,tree}(\alpha\z),
\nn \\
\bm{B} && = 4a_sC_F \Gamma(1-\e)B^\e \int d\alpha \alpha\bar{\alpha} \widehat{h}_1^\perp(\alpha\zeta),
\nn\\
\bm{C+C^*} &&=  
  2a_s \Gamma(-\e)\( C_F-\frac{C_A}{2}\) B^\e \int_0^1 d\alpha \int_L^{-L} d\tau    \( \alpha  \widehat{E}(\tau,\z_{\z\tau}^\alpha,0) -\alpha\widehat{E}(\alpha \zeta, \tau,0) \),
\nn \\
\bm{D+D^*} &&= 2a_s \Gamma(-\e) B^\e \frac{C_A}{2}\int [d\alpha d\beta d\gamma] \int_L^{-L} d\tau \[ -\bar{\beta} \widehat{E}(\z_{\z\tau}^\alpha,\z_{\tau\z}^\beta,0) -\frac{\bar{\alpha}\bar{\beta}}{\alpha}\widehat{E}(\z_{\z\tau}^\alpha,\z_{\tau\z}^\beta,0) - \widehat{E}(\bar{\alpha} \z,\z_{\tau\z}^\beta , 0)\],
\nn \\
\bm{E+E^*} &&= 2 a_s\( C_F-\frac{C_A}{2}\) \Gamma(1-\e)B^\e \int_0^1 d\alpha  \ \alpha\bar{\alpha} \int_L^{-L}d\tau  \widehat{E}(\alpha \z; \tau;0),
\nn \\
\bm{F} &&= 2a_s \frac{C_A}{2} \Gamma(1-\e)B^\e \int_0^1 d\alpha \ \alpha \bar{\alpha} \int_L^{-L} d\tau  \widehat{E}(\alpha \z,\tau,0).
\end{eqnarray}
It is straightforward to convince oneself that the sum $\bm{B+E+E^*+F}$ vanishes identically. Therefore, the only non-zero contribution is to the pole part.

\subsection{Worm-gear-L function $h_{1L}^\perp$}
\label{app:diagrams:wgh}
The worm-gear function $h_{1L}^\perp$ behaves similar to the worm-gear function $g_{1T}$, but has no gluon-contributions. Specifically, one has both twist-three and twist-three tree-level matching:
\begin{align}
\widehat{h}_{1L}^{\perp,tree}(\zeta) &= \frac{1}{i\zeta}\( 2\int_0^1d\alpha \alpha \widehat{h}_1(\alpha\zeta) - \widehat{h}_1(\zeta) \) + i\zeta \int_0^1d\beta \beta \widehat{H}(\zeta,\beta\zeta,0) \notag\\
 & -i\zeta \int [d\alpha d\beta d\gamma]  2\beta^2 \widehat{H}(\bar{\alpha}\zeta,\beta\zeta,0)-\frac{i}{2} \(\int_L^0d\tau-\int_0^{-L}d\tau\)  \widehat{H}(\zeta,\tau,0),\notag \\
 \widehat{h}_L(\zeta) &= 2\int_0^1d\alpha \alpha \widehat{h}_1(\alpha\zeta)+ 2 \zeta^2 \int [d\alpha d\beta d\gamma] \beta^2 \widehat{H}(\bar{\alpha}\zeta,\beta\zeta,0),\label{app:hL}
\end{align}
where the function $\widehat{H}$ obeys the symmetry relation
\begin{equation}
\widehat{H}(\zeta_1,\zeta_2,\zeta_3) = -\widehat{H}(-\zeta_3,-\zeta_2,-\zeta_1).
\end{equation}

For individual diagrams, we find:
\begin{eqnarray}
\bm{A+A^*} &&=  2a_s C_F B^\e \Gamma(-\e) \int_0^1 d\alpha \(\frac{2\alpha}{\bar{\alpha}}\)_+ \alpha \widehat{h}_{1L}^{\perp,tree}(\alpha\zeta) - 2\delta(\bar{\alpha})(1+\lambda_\delta)\widehat{h}_{1L}^{\perp,tree}(\alpha\zeta) ,
\nn \\
\bm{B} &&= -2a_sC_F \Gamma(1-\e)B^\e \int d\alpha \ 2\alpha\bar{\alpha} \widehat{h}_{1L}^{\perp,tree}(\alpha\zeta) -(1-2\alpha) \frac{\widehat{h}_L(\alpha\zeta)}{i\zeta},
\nn \\
\bm{C+C^*} &&= -ia_s \Gamma(-\e)\( C_F-\frac{C_A}{2}\) B^\e p^+ \int [d\alpha  d\beta d\gamma] \[\int_L^{0} d\tau - \int_0^{-L} d\tau \] 
\nn \\
&& \times \[ 2\frac{\beta}{\bar{\beta}} \widehat{H}(\beta \z,\tau-\alpha  \z, 0) + 2\frac{\gamma}{\bar{\beta}} \widehat{H}(\z_{\tau\z}^\beta,\alpha (\tau-\z),0) - 2\e  \widehat{H}(\z_{\tau\z}^\beta,\alpha (\tau-\z),0 )\],
\nn \\
\bm{D+D^*} &&= i a_s\frac{C_A}{2}B^\e \Gamma(-\e)\int [d\alpha d\beta d\gamma] \( \int_L^{0}d\tau-\int_0^{-L}d\tau\)  \Bigg\{
\nn \\
&& \( \frac{2\bar{\alpha}}{\beta}-2\e\) \widehat{H}(\zeta_{\z\tau}^\beta,\gamma(\z-\tau),0)+ \frac{2\bar{\beta}}{\beta}\widehat{H}(\bar{\beta}\z,\tau+\alpha \z,0) \Bigg\},
\nn \\
\bm{E+E^*} &&= 2ia_s\( C_F-\frac{C_A}{2}\) \Gamma(1-\e)B^\e 
\\\nn && \times \int [d\alpha d\beta d\gamma] \(\int_L^{0}d\tau-\int_0^{-L}d\tau\)  (2\gamma -1 + \e) \widehat{H}(\alpha \z, \tau -\beta \z,0),
\nn \\
\bm{F} &&= -2i \Gamma(1-\e)a_s \frac{C_A}{2}B^\e \int [d\alpha d\beta d\gamma] \ \beta \( \int_L^0 d\tau-\int_0^{-L}d\tau\) \widehat{H}(\bar{\beta} \z,\tau+ \alpha \z,0).
\end{eqnarray}
It is interesting to observe that after substitution of (\ref{app:hL}) the twist-two part of the diagram $\mathbf{B}$ vanishes. It leads to a trivial finite part for the twist-two contribution.


\providecommand{\href}[2]{#2}\begingroup\raggedright\endgroup

\end{document}